\documentclass[a4paper,floatfix,accepted=2026-02-13]{quantumarticle}
\pdfoutput=1
\usepackage{graphicx} 
\usepackage{pgfplots}
\pgfplotsset{compat=1.18}
\usepackage[normalem]{ulem}
\usepackage{amsmath}
\usepackage{amssymb}
\usepackage{amsthm}
\usepackage{physics}
\usepackage{comment}
\usepackage{tikz}
\usetikzlibrary{decorations.pathreplacing}
\usepackage{quantikz}
\usepackage{xcolor}
\usepackage{hyperref}
\usepackage[capitalise]{cleveref}
\usepackage[numbers]{natbib}

\newcommand{\diag}{\text{diag}}
\newcommand{\calU}{\mathcal{U}}

\newcommand{\poly}{\mathrm{poly}}
\newcommand{\Hi}{H_{\mathrm{init}}}
\newcommand{\Hf}{H_{\mathrm{final}}}
\newcommand{\sigmai}{\sigma_{\mathrm{init}}}
\newcommand{\sigmaf}{\sigma_{\mathrm{final}}}
\newcommand{\Tf}{T_{\mathrm{filter}}}

\newcommand{\calZ}{\mathcal{Z}}
\newcommand{\Eqate}{E_{\mathrm{QATE}}}
\newcommand{\dEqate}{\Delta E_{\mathrm{QATE}}}
\newcommand{\Em}{E_{\mathrm{min}}}
\newcommand{\Eg}{E_{\mathrm{G}}}
\newcommand{\dEmin}{\Delta E_{\mathrm{min}}}
\newcommand{\Var}{\mathrm{Var}}
\newcommand{\qate}{\mathrm{QATE}}

\newcommand{\Zi}{\mathcal{Z}_{\mathrm{init}}}
\newcommand{\Zf}{\mathcal{Z}_{\mathrm{final}}}
\newcommand{\rhoi}{\rho_{\mathrm{init}}}
\newcommand{\rhoadi}{\rho_{\mathrm{QATE}}}
\newcommand{\rhomin}{\rho_{\mathrm{min}}}

\definecolor{MLM}{cmyk}{0.1,0.8,0,0.1} 
\definecolor{MDG}{rgb}{0,0.55,0.05} 
\definecolor{atomictangerine}{rgb}{1.0, 0.6, 0.4}
\definecolor{bluegray}{rgb}{0.4, 0.6, 0.8}
\definecolor{brightube}{rgb}{0.82, 0.62, 0.91}
\definecolor{brilliantlavender}{rgb}{0.96, 0.73, 1.0}

\begin{document}
\title{Quasi-Adiabatic Processing of Thermal States}
\date{May 26, 2025}
\author{Reinis Irmejs}
\email{reinis.irmejs@mpq.mpg.de}
\affiliation{Max-Planck-Institut für Quantenoptik, Hans-Kopfermann-Straße 1, D-85748 Garching, Germany}
\affiliation{Munich Center for Quantum Science and Technology (MCQST), Schellingstraße 4, D-80799 Munich, Germany}

\author{Mari Carmen Bañuls}
\affiliation{Max-Planck-Institut für Quantenoptik, Hans-Kopfermann-Straße 1, D-85748 Garching, Germany}
\affiliation{Munich Center for Quantum Science and Technology (MCQST), Schellingstraße 4, D-80799 Munich, Germany}

\author{J. Ignacio Cirac}
\affiliation{Max-Planck-Institut für Quantenoptik, Hans-Kopfermann-Straße 1, D-85748 Garching, Germany}
\affiliation{Munich Center for Quantum Science and Technology (MCQST), Schellingstraße 4, D-80799 Munich, Germany}

\begin{abstract}
We investigate the performance of an adiabatic evolution protocol when initialized from a Gibbs state at finite temperature. Specifically, we identify the diagonality of the final state in the energy eigenbasis, as well as the difference in energy and in energy variance with respect to the ideal adiabatic limit as key benchmarks for success and introduce metrics to quantify the off-diagonal contributions. Provided these benchmarks converge to their ideal adiabatic values, we argue that thermal expectation values of observables can be recovered, in accordance with the eigenstate thermalization hypothesis. For the transverse-field Ising model, we analytically establish that these benchmarks converge polynomially in both the quasi-adiabatic evolution time $T$ and system size. We perform numerical studies on non-integrable systems and find close quantitative agreement for the off-diagonality metrics, along with qualitatively similar behavior in the energy convergence.
\end{abstract}

\maketitle

\section{Introduction}\label{sec:introduction}

Controllable quantum devices provide a natural approach to investigate physical quantum systems. Recent advances in platforms using superconducting circuits \cite{ai2024quantum}, trapped ions \cite{liu2025certified}, Rydberg atoms \cite{bluvstein2024logical}, and cold atoms \cite{bloch2017} offer unprecedented control to probe quantum many-body systems, directly avoiding the exponential overhead incurred when simulating these systems on classical computers.

A particularly important approach in quantum simulation is the quantum adiabatic algorithm (QAA), which allows for the preparation of ground states through time evolution. In this approach, the system is initialized in the ground state of a trivial Hamiltonian and gradually evolved to the ground state of a target Hamiltonian by interpolating between the two. The success of this method relies on the existence of a continuous interpolation path with a non-vanishing spectral gap and a sufficiently slow transition throughout. While finding the ground state of a generic quantum system is computationally hard in general \cite{LocalHComp}, the QAA provides a powerful strategy for systems with a spectral gap \cite{born1928beweis,Kato1950,Jansen2007, Lidar2018}. Since the algorithm only requires time evolution, it offers a particularly accessible strategy for near-term quantum computation and quantum simulation platforms.

It is thus natural to ask whether the QAA can be extended beyond ground states to prepare excited eigenstates, and thereby approximate Gibbs states, which are statistical mixtures of all eigenstates. In principle, this is possible; however, for typical many-body systems, the excited states in the bulk of the spectrum are exponentially close in system size, meaning the gaps between neighboring levels become exponentially small. As a result, maintaining adiabaticity requires exponentially long runtimes \cite{Illin2021}. Furthermore, even if the adiabaticity is ensured, the resulting state will not necessarily be the Gibbs ensemble, unless the spectra of initial and final Hamiltonians are exactly proportional to each other. 

Application of the adiabatic protocol to Gibbs states has been studied in the low-temperature regime. In \cite{greenblatt2022}, the authors investigate the low-temperature regime in the presence of a weakly interacting perturbation, showing that local observables remain close to their thermal expectation values. In contrast, \cite{zuo2023} finds that adiabatically evolving from a non-interacting to an interacting Hamiltonian does not, in general, reproduce the target Gibbs state; however, in the adiabatic limit, the resulting state remains the closest approximation to it. The special case, where the spectra of the initial and final Hamiltonians are proportional, was studied in \cite{Plastina2014, Plastina2019}. Studies of the adiabatic evolution of excited quantum scar states reveal a similar behavior to ground state adiabatic dynamics \cite{Yarloo2024}. An analysis in \cite{Carcy_2021} investigated the use of the QAA in ultracold atom platforms, where the initial state is not necessarily at zero temperature \cite{bloch2008many, trotzky2010suppression}.  
Despite these advances, a complete understanding of when and why adiabatic protocols succeed at finite temperature remains an open question.

The strict conditions required for exact adiabaticity are, in general, too restrictive, since the gaps between neighboring eigenvalues are exponentially small in the system size. However, we argue that, to obtain accurate expectation values of thermal observables, some mixing of eigenvectors may be tolerated. Indeed, according to the eigenstate thermalization hypothesis (ETH), thermal observables can still be reproduced in the thermodynamic limit as long as the off-diagonal elements remain small and the energy fluctuations remain at most linear in the system size. This motivates us to investigate to what extent a thermal state evolved under a slowly varying Hamiltonian retains thermal properties with respect to the final Hamiltonian. More concretely, we define a quasi-adiabatic thermal evolution (QATE) as a process in which a unitary evolution—formally equivalent to that in the QAA—is applied to an initial thermal state. We focus on the practically relevant regime where the total evolution time does not scale exponentially with system size, thereby breaking strict adiabaticity. In contrast to previous works, we carry out our analysis across the full temperature range, with an emphasis on assessing worst-case performance.

To assess the extent of the eigenstate mixing and adiabaticity breaking, we focus on investigating diagonality of the final state and introduce benchmarks to quantify it. Furthermore, this evolution can also be interpreted as an isentropic cooling process, where the energy of the final state gradually decreases to the ideal adiabatic limit, while preserving entropy (see \cref{fig:approach}). To quantify how closely the evolution approaches the ideal adiabatic limit, we investigate the deviation in final energy from the ideal case, denoted $\dEqate$, providing an additional benchmark for adiabaticity breaking. Together with a variance that scales consistently with a Gibbs state, a small enough deviation in energy ensures that local observables will be sufficiently thermal (in the generic case).

The QATE protocol is conceptually related to thermofield double (TFD) based approaches to thermal state preparation, in which one prepares a purification of a Gibbs state by introducing an auxiliary copy of the system \cite{ISRAEL1976, Juan_Maldacena_2003, ChapmanTFD2019}. In this framework, one engineers a doubled-system Hamiltonian whose ground state is (approximately) the TFD and prepares this ground state, for instance via adiabatic evolution. While powerful, this strategy comes with important overheads: it doubles the Hilbert space and requires constructing an appropriate TFD parent Hamiltonian, which is nontrivial for interacting models and can involve nonlocal couplings in the doubled system \cite{CottrellTFD2019, Failde2025}. In particular, even if the target Gibbs state corresponds to a local physical Hamiltonian, the engineered Hamiltonian used to prepare its purification need not inherit this locality in a straightforward manner.
In contrast, QATE does not aim to prepare an exact Gibbs state and does not require reproducing precise Gibbs weights. Instead, it is sufficient that the evolved state becomes sufficiently diagonal in the eigenbasis of the target Hamiltonian, which already enables access to thermal expectation values. Moreover, QATE does not require enlarging the system size and, when implemented via interpolation between local Hamiltonians, can be realized using local, time-dependent dynamics. QATE also has distinct experimental relevance in ultracold atom platforms, where adiabatic evolution is routinely employed starting from finite-entropy states corresponding to finite temperatures \cite{bloch2008many, trotzky2010suppression}. Nevertheless, such approaches have proven highly successful in investigating thermal and equilibrium phenomena, including the Mott transition in bosonic \cite{greiner2002quantum} and fermionic \cite{Esslinger2010} Hubbard models.

The rest of this manuscript is structured as follows. In \cref{sec:methods}, we review the adiabatic theorem and formally introduce the quasi-adiabatic protocol. \cref{sec:results} presents analytical results for the transverse field Ising model, showing a $\poly(N,T^{-1})$ convergence to the ideal adiabatic limit. Additionally, we numerically investigate both free fermionic and non-integrable systems, reproducing the same convergence behavior. We discuss our findings and finalize the work by final remarks in \cref{sec:conclusions}.

\definecolor{maroon}{RGB}{178, 34, 34}
\definecolor{navyblue}{RGB}{30, 60, 150}
\definecolor{pinegreen}{RGB}{1,121,111}
\begin{figure}
    \centering
    \begin{minipage}{\linewidth}
    \centering
    \begin{tikzpicture}
        \begin{axis}[
            axis lines = middle,
            xlabel={$S[\rho]$},
            ylabel={$\Tr(\rho \Hf)$},
            samples=100,
            domain=0.001:0.5,
            enlargelimits=true,
            width=10cm, 
            height=8cm, 
            xtick=\empty, 
            ytick=\empty, 
            xtick style={black}, 
            ytick style={black}  
        ]
            \pgfmathsetmacro{\c}{0.45}
    
            \addplot[navyblue, thick] 
            ({-x*ln(x)-(1-x)*ln(1-x)}, {2*x-1}) node[right] {};
            
            \addplot[maroon, thick] 
            ({-x*ln(x)-(1-x)*ln(1-x)}, {\c*(2*x-1)}) node[right] {};
            
            \pgfmathsetmacro{\calt}{0.05}
            \pgfmathsetmacro{\xval}{-\calt*ln(\calt) - (1-\calt)*ln(1-\calt)} 
            \pgfmathsetmacro{\yval}{\c*(2*\calt-1)}  
            \pgfmathsetmacro{\yvalshifted}{\yval - 0.35}  
            \pgfmathsetmacro{\yvalG}{(2*\calt-1)}  
            
            \addplot[only marks, mark=*, mark size=3pt, maroon] coordinates {(\xval,\yval)};
            \addplot[only marks, mark=*, mark size=3pt, black] coordinates {(\xval,\yvalshifted)};
            \addplot[only marks, mark=*, mark size=3pt, pinegreen] coordinates {(\xval,\yvalshifted-0.1)};
            \addplot[only marks, mark=*, mark size=3pt, navyblue] coordinates {(\xval,\yvalG)};
            \addplot[only marks, mark=square*, mark size=3pt, navyblue] coordinates {(\xval+0.172,\yvalshifted)};

            \addplot[only marks, mark=*, mark size=1pt, black] coordinates {(0,\yvalshifted)};
            \addplot[only marks, mark=*, mark size=1pt, pinegreen] coordinates {(0,\yvalshifted-0.1)};
            \addplot[only marks, mark=*, mark size=1pt, navyblue] coordinates {(0,\yvalG)};
            
            \addplot[gray, dashed, opacity=0.5] coordinates {(\xval,-1) (\xval,0)};
    
            \addplot[navyblue, dashed, opacity=0.5] coordinates {(0,\yvalG) (\xval+0.22,
            \yvalG)};
            \addplot[black, dashed, opacity=0.5] coordinates {(0,\yvalshifted) (\xval+0.22,
            \yvalshifted)};
            \addplot[pinegreen, dashed, opacity=0.5] coordinates {(0,\yvalshifted-0.1) (\xval+0.22,
            \yvalshifted-0.1)};
            
            \draw[->, thick] (axis cs:\xval,\yval) -- (axis cs:\xval,\yvalshifted+0.01);
            
            \node[right, maroon] at (axis cs:\xval,\yval-0.05) {$\rhoi$};
            \node[right, black] at (axis cs:\xval,\yvalshifted+0.03) {$\rhoadi$};
            \node[left, pinegreen] at (axis cs:\xval,\yvalshifted-0.07) {$\rhomin$};
            \node[right, navyblue] at (axis cs:\xval,\yvalG-0.02) {$\rho_{\mathrm{G}}$};
            \node[right, navyblue] at (axis cs:\xval+0.15,\yvalshifted+0.05) {$\rho_{\mathrm{E}}$};

            \node[right, black] at (axis cs:0,\yvalshifted+0.03) {$\Eqate$};
            \node[right, pinegreen] at (axis cs:0,\yvalshifted-0.07) {$\Em$};
            \node[right, navyblue] at (axis cs:0,\yvalG-0.03) {$\Eg$};
            
            \node[above, navyblue,rotate=48] at (axis cs:\xval+0.4, {\yval}) {$\exp(-\beta \Hf)$};
            \node[above, maroon,rotate=18] at (axis cs:\xval+0.25, {\yval+0.1}) {$\exp(-\beta \Hi)$};
    
            \draw[decorate, decoration={brace, mirror,amplitude=5pt}, thick]
            (axis cs:\xval+0.22, \yvalshifted-0.1) -- (axis cs:\xval+0.22, \yvalshifted)
            node[midway, right=4pt] {$\dEqate$};
    
            \draw[decorate, decoration={brace,mirror, amplitude=5pt}, thick]
            (axis cs:\xval+0.22, \yvalG) -- (axis cs:\xval+0.22, \yvalshifted-0.1)
            node[midway, right=4pt] {$\dEmin$};

            \node[below] at (axis cs:0.08,-0.05) {$\beta = \infty$};
    
    
            \node[below] at (axis cs:0.60,-0.05) {$\beta = 0$};
            
        \end{axis}
    \end{tikzpicture}
    \end{minipage}
    \caption{Graphical illustration of the QATE. The Gibbs state minimizes the energy at fixed von Neumann entropy $S[\rho] = -\Tr(\rho \ln(\rho))$, so the curve of initial states (red) lies above that of final states (blue). The adiabatic evolution is unitary and conserves the von Neumann entropy, corresponding to vertical motion in the diagram. This corresponds to isentropic cooling. While $\Eg$ represents the minimal energy at a given entropy, the preservation of the initial spectrum constrains the minimum attainable energy to $\Em$, given by state $\rhomin$. Running the protocol for a finite time will prepare $\rhoadi$ with energy $\Eqate\geq\Em$. We denote $\rho_{\mathrm{E}}$ and $\rho_{\mathrm{G}}$ as the Gibbs states at the same energy and entropy as $\rhoadi$, respectively.
    }
    \label{fig:approach}
\end{figure}
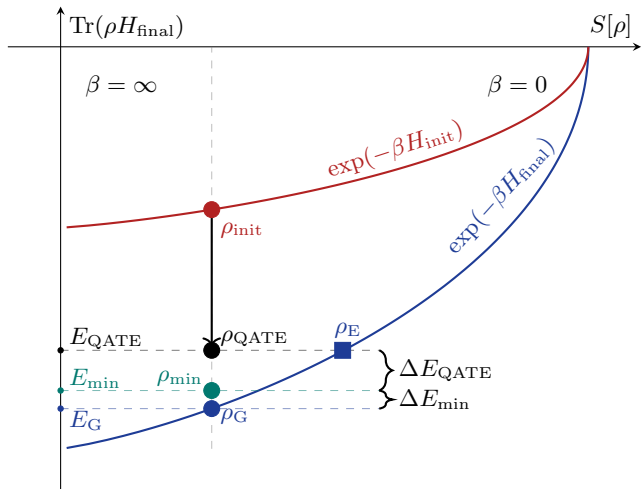

\section{Methods}\label{sec:methods}

\subsection{The adiabatic theorem}\label{sec:QAA}

We begin by summarizing QAA for the preparation of ground states. The QAA directly relies on the adiabatic theorem, which states that an evolving system remains in the instantaneous eigenstate if the system Hamiltonian is varied sufficiently slowly and the eigenstate is separated from the other eigenstates by a non-vanishing gap \cite{messiah_quantum_2020}. Formally, for an initial Hamiltonian $\Hi$ with a ground state $\ket{\Psi_{\mathrm{init}}}$ and a desired target Hamiltonian $\Hf$ the instantaneous Hamiltonian $H(s)$ is given by
\begin{equation}\label{eq:Ham_adi}
    H(s) = [1-\gamma(s)]\Hi + \gamma(s)\Hf,
\end{equation}
where $\gamma(s)$ is the adiabatic schedule such that $\gamma(0)=0$ and $\gamma(1)=1$ and $s=t/T$ is the parametrized time. The QAA proceeds as
\begin{align}
    \ket{\Psi_{\mathrm{adi}}(T)} &= \calU \ket{\Psi_{\mathrm{init}}},\\
    \label{eq:unitaryQAA}
    \calU = \mathcal{T} &e^{-i\int_0^T H\left(\frac{t}{T}\right) dt},
\end{align}
where $\mathcal{T}\exp(\cdot)$ denotes the time ordered exponential. The minimum adiabatic runtime $T$ required to achieve a precison $\epsilon$ such that $\abs{\bra{\Psi_{\mathrm{final}}}\ket{\Psi_{\mathrm{adi}}(T)}}^2=(1-\epsilon)$ depends closely on the spectral gap $\Delta(t)$ throughout the evolution as
\begin{equation}
\label{eq:qaa}
    T=\mathcal{O} \left (\max_s \frac{\norm{\partial_s H(s)}+\norm{\partial^2_s H(s)}}{\epsilon^{-1/2}\Delta^2(s)}+ \frac{\norm{\partial_s H(s)}^2}{\epsilon^{-1/2}\Delta^3(s)}\right),
\end{equation}
where $\norm{\cdot}$ denotes the operator norm \cite{Jansen2007, Amin2009, Lidar2018}. 
In particular, 
for a linear ramp $\gamma(s)=s$ and fixed system size (and system size independent gap) the result implies that:
\begin{equation}\label{eq:fixed_N_dep}
    \epsilon = \mathcal{O}\left (\frac{1}{T^2}\right).
\end{equation}
The results of \cite{nenciu1993linear, hagedorn2002elementary, Lidar2018} show that the dependence on the error can be further improved if the derivatives of $\gamma(s)$ vanish for $s=0,1$. If sufficiently many derivatives vanish, one can achieve error that scales as $\mathcal{O}(e^{-cT})$. 

\subsection{Quasi-adiabatic thermal evolution}\label{sec:QATE}

Drawing inspiration from QAA, we introduce a similar protocol 
for processing Gibbs states,
with the goal to probe the thermal properties of the system described by a certain Hamiltonian $\Hf$ at a finite temperature.
The idea is to start from an easily preparable Gibbs state of a different Hamiltonian $\Hi$, 
\begin{equation}
    \rhoi = \frac{e^{-\beta \Hi}}{\Zi}, \quad \text{where} \quad \Zi = \Tr(e^{-\beta \Hi}),
\end{equation}
(we consider $\beta>0$) and to evolve it using the same unitary evolution $\calU$ employed in QAA [see \cref{eq:unitaryQAA}].

As discussed in the introduction, eigenstates in the bulk of the spectrum are separated by energy gaps that shrink exponentially with system size. This implies that true adiabaticity can only be maintained if the runtime $T$ scales exponentially with the system size $N$. Here, we focus on the regime of non-exponential runtimes, where adiabaticity is only approximately preserved. We refer to this approach as the Quasi-Adiabatic Thermal Evolution (QATE).
The resulting state of this protocol will thus be denoted
\begin{equation}
    \rhoadi(T) = \calU \rhoi \calU^\dagger.
\end{equation}

While in QAA one can directly compare the final state to the ground state of $\Hf$, in QATE such a direct comparison is not straightforward. Importantly, since QATE consists of a unitary evolution, it conserves the entropy of the initial mixed state and preserves the eigenvalues of $\rhoi$ (see \cref{fig:approach}). In \cref{sec:benchmarking}, we discuss how to benchmark the quality of $\rhoadi(T)$, taking into account these conserved quantities.

\subsection{Benchmarking QATE}\label{sec:benchmarking}

Firstly, an important benchmark for quantifying the extent of adiabaticity breaking is the energy of QATE state $\rhoadi(T)$ with respect to $\Hf$. Since QATE is a unitary protocol, it preserves both the von Neumann entropy $S[\rho] = -\Tr(\rho \ln(\rho))$ and the eigenvalues of the initial state $\rhoi$. The conservation of the spectrum imposes a fundamental limit on the minimal attainable energy.
Consider an initial Hamiltonian $\Hi = \sum_k E^i_k \ket{E^i_k}\bra{E^i_k}$ with
\begin{equation}
    \rhoi = \frac{1}{\Zi}\sum_k e^{-\beta E^i_k}\ket{E^i_k}\bra{E^i_k},
\end{equation}
and a final Hamiltonian $\Hf = \sum_k E^f_k \ket{E^f_k}\bra{E^f_k}$. The state with the lowest possible energy, while maintaining the same spectrum as $\rhoi$, is the diagonal state
\begin{equation}\label{eq:rhomin}
    \rho_{\mathrm{min}} = \frac{1}{\Zi}\sum_k e^{-\beta E^i_k}\ket{E^f_k}\bra{E^f_k},
\end{equation}
where both the energies $E^i_k$ and $E^f_k$ are ordered increasingly. The corresponding minimal energy is given by
\begin{equation}
    E_{\mathrm{min}} = \Tr(\rho_{\mathrm{min}} \Hf) = \frac{1}{\Zi}\sum_k e^{-\beta E^i_k} E^f_k.
    \label{eq:emin}
\end{equation}
Importantly, $\rho_{\mathrm{min}}$ also coincides with the state obtained in the limit of exponentially long runtimes, as perfect adiabaticity prevents eigenstate mixing and preserves diagonality in the instantaneous (and thus final) eigenbasis, assuming non-degenerate spectrum.

This motivates us to view QATE as a form of isentropic cooling. Since the energy is minimized in the limit $T \to \infty$, we expect the energy of QATE state to approach this value as the runtime $T$ increases. We denote the energy distance between QATE state at time $T$, $\rhoadi(T)$, and $\rho_{\mathrm{min}}$ by $\dEqate$. A central aim of this work is to study how $\dEqate$ decreases with $T$ and converges toward the limit of true adiabaticity. Since the Gibbs state minimizes the energy for a given von Neumann entropy, isentropic cooling in QATE also implies that the system approaches the Gibbs state energy.

However, in general, it will not be possible to reach the Gibbs state. In particular, if the spectra of $\Hi$ and $\Hf$ are not exactly proportional to each other (up to a constant shift), $\Em$ in \cref{eq:emin} will not coincide with the energy of a Gibbs state of $\Hf$, imposing a fundamental limitation that prevents exact convergence. We quantify the size of this barrier by the energy difference $\dEmin = \Em - \Eg$, where $\Eg$ is the energy of the true Gibbs state of $\Hf$ at the same von Neumann entropy. A graphical illustration of this process is provided in \cref{fig:approach}.

Provided that the state $\rhomin$ has a sufficiently narrow energy width, it guarantees that the thermal values of physical observables can be obtained in the generic (ETH) case~\cite{Rigol_2008, Rigol_2016}. 
However, when the adiabatic condition is broken, the evolution $\calU$ will inevitably induce off-diagonal matrix elements, and we will obtain a state of the general form
\begin{equation}
    \rhoadi = \sum_{ij} c_{ij} \ket{E^f_i}\bra{E^f_j}.
\end{equation}
We are thus interested in how large the off-diagonality is, depending on the total evolution time, and how the magnitude of the off-diagonal coefficients depends on the energy difference $\abs{E^f_i - E^f_j}$.
Examining the coefficients $c_{ij}$ individually is not feasible due to the exponential size of the Hilbert space. Instead, we probe how the off-diagonal coefficients collectively depend on the difference between their eigenstate energies through the \emph{binned off-diagonality} (BOD)
\begin{align}
    \mathrm{BOD}=\frac{\sum_{\abs{E_i - E_j} \in \{\omega-\delta, \omega+\delta\}} \abs{c_{ij}}^2}{\Tr(\rhoadi^2)},
\end{align}
where the sum is taken over a finite energy window of width $2\delta$. The window $\delta$ is chosen to balance statistical sampling and resolution, being large enough to include many eigenstate energy differences while remaining small enough to resolve the dependence of the BOD on the energy difference $\omega$.
We normalize by the purity $\Tr(\rhoadi^2)$, to ensure $0\leq \mathrm{BOD}\leq 1$ and allow for comparison between different system sizes. 
The BOD can be computed directly from exact diagonalization (ED) calculations or, approximately, using filter techniques~\cite{Lu_2021}. In the latter case, instead of a top-hat energy window of width $2\delta$, a Gaussian of a standard deviation $\delta$ is used (see \cref{app:filter} for the derivation and more details).

A simpler quantity that probes off-diagonality is the \emph{commutator off-diagonality} (COD), defined as
\begin{align}
    \mathrm{COD}=-\frac{\Tr\left(\comm{\Hf}{\rhoadi}^2\right)}{\Tr(\rhoadi^2)} = \frac{\sum_{ij} \abs{c_{ij}}^2 (E^f_i - E^f_j)^2}{\sum_{ij} \abs{c_{ij}}^2}.
\end{align}
COD assigns a single value characterizing how off-diagonal a given state is. Like BOD, COD is normalized by the purity. However, it retains a system size $N$ dependence that we quantify in the next section. 

Both BOD and COD quantify the deviation from true adiabaticity, since in the exponential-$T$ limit the resulting state $\rhomin$ becomes fully diagonal. Thus, BOD and COD are key properties for assessing the quality of QATE. Additionally, both COD and BOD can be easily evaluated numerically.

Thirdly, we aim for a $\rhoadi$ with an energy variance that scales linearly with the system size $N$, with respect to the final Hamiltonian,
\begin{equation}
\Var(\rho)=\Tr(\rho \Hf^2) - \Tr(\rho \Hf)^2 .
\end{equation}
Provided that the final state is diagonal and exhibits a linearly scaling energy variance $\Var(\rho_{\mathrm{adi}})=O(N)$, the ETH predicts that its properties are indistinguishable from those of a Gibbs state at the same mean energy \cite{Rigol_2008, Rigol_2016}. We expect that along with COD and $\dEqate$, the variance will converge to the value of the ideal adiabatic evolution as $T$ increases. To investigate this, we compare the variance of $\rhoadi$ with the variance of $\rhomin$ [\cref{eq:rhomin}]
\begin{equation}
    \Delta \Var_{\qate} = \Var(\rhoadi) - \Var(\rhomin),
\end{equation}
as well as the relative error $\Delta \Var_{\qate}/\Var_{\qate}$, to verify that the scaling is extensive. 

Notably, QATE reduces to QAA in the zero-temperature limit $\beta = \infty$, where provable performance guarantees exist, as discussed earlier in \cref{sec:QAA}. Similarly, in the infinite-temperature limit $\beta = 0$, both the initial and final states are proportional to the identity, leading to trivial success. In \cref{app:gaussian}, we investigate the case where the density of states of both the initial and final Hamiltonians is Gaussian—a good approximation for generic local Hamiltonians in the small-$\beta$, large-$N$ limit. In this regime, we find that $\dEmin = 0$, and $\Var(\rhomin) = \Var(\rho_{\mathrm{G}})$. When benchmarking QATE, we will avoid these limiting cases and instead focus on the most challenging behavior at moderate values of $\beta$.

\section{Results and Discussion}\label{sec:results}

In this section, we present both analytical and numerical results on quantum spin systems. For the numerical simulations, we use a Trotterized \cite{Childs_2021} version of QATE,
\begin{equation}
    \calU = \mathcal{T} e^{-i\int_0^T H\left(\frac{t}{T}\right) dt} \approx \prod_{j=1}^M e^{-iH\left(\frac{j}{M}\right)\tau},
\end{equation}
where the time step is set to $\tau = 0.1$ throughout. 
We adopt a linear ramp $\gamma(s) = s$ for simplicity but include results for a smoothed evolution in \cref{app:smooth_ramp}. As discussed in \cref{sec:benchmarking}, we avoid the (trivial) extreme temperatures and focus on a moderate value $\beta = 1$.

\subsection{Translationally Invariant Free Fermions}\label{sec:tfi_ti_results}

In this subsection, we present both analytical and numerical results for the transverse field Ising model (TFIM). We choose this model for its simplicity and the possibility of performing analytical calculations for QATE. Although the model is fully integrable and thus not generic, it still provides valuable insights into how the benchmarks of energy, diagonality, and variance behave. Furthermore, we will use the analytical results obtained here as a reference for comparisons with more complex systems.

The Hamiltonian reads
\begin{equation}
    \label{eq:TFI}
    H_{\mathrm{TFI}} = -J\sum_{n=1}^{N-1}\sigma_n^x \sigma_{n+1}^x - g \sum_{n=1}^{N} \sigma_n^z - J_{bc} \sigma_N^x \sigma_1^x,
\end{equation}
where $J_{bc}$ encodes the boundary conditions, and $\sigma_i^{\alpha}, \;\alpha = x,y,z$ are Pauli operators acting on site $i$. We set $J = 1$ and choose $J_{bc} = -P$, where $P = \prod_{i=1}^N \sigma^z_i$ is the parity operator, resulting in anti-periodic boundary conditions in the even subspace and periodic boundary conditions in the odd subspace. This choice corresponds to a translationally invariant model in the free-fermion picture 
that simplifies the analytical treatment.

The mapping of $H_{\mathrm{TFI}}$ to a free fermion model and its diagonalization are explicitly shown in \cref{app:tfi-exact}.
In the momentum space, the modes with momenta $\pm k$ couple only to each other (for $k\neq 0, N/2$), confining the dynamics to these subsystems. 
The model can then be easily diagonalized via a Bogoliubov transformation. Furthermore, 
because of the symmetry under fermion parity, within each $(k,-k)$ block the dynamics splits in two disconnected subsectors,   
such that the problem reduces to a collection of two-level systems. In each of these blocks, the off-diagonality is quantified by a single coefficient $c_k$, that we can bound using the results from QAA. All of the key benchmarks COD, $\dEqate$ and $\Delta \mathrm{Var}_{\mathrm{QATE}}$ can be expressed through the off-diagonal coefficients $c_k$, which allows us to determine their scaling.

Notice that even though the model has  exponentially small gaps between eigenstates, the evolution does not couple such states directly. During QATE, the instantaneous Hamiltonian only connects eigenstates that differ in the occupation of a single $(k,-k)$ block. Away from the critical point (and except for the lowest mode), such states are separated by at least a $1/\poly(N)$ energy difference, which therefore sets the relevant gap for the evolution. Exponentially small gaps, by contrast, occur only between states that differ in the occupation of multiple modes and are not directly coupled by the dynamics. At the critical point, however, the lowest-mode gap closes, and the zero-temperature phase transition also becomes apparent at finite temperatures.

\begin{figure*}[t!]
    \centering
    \includegraphics[width=\linewidth]{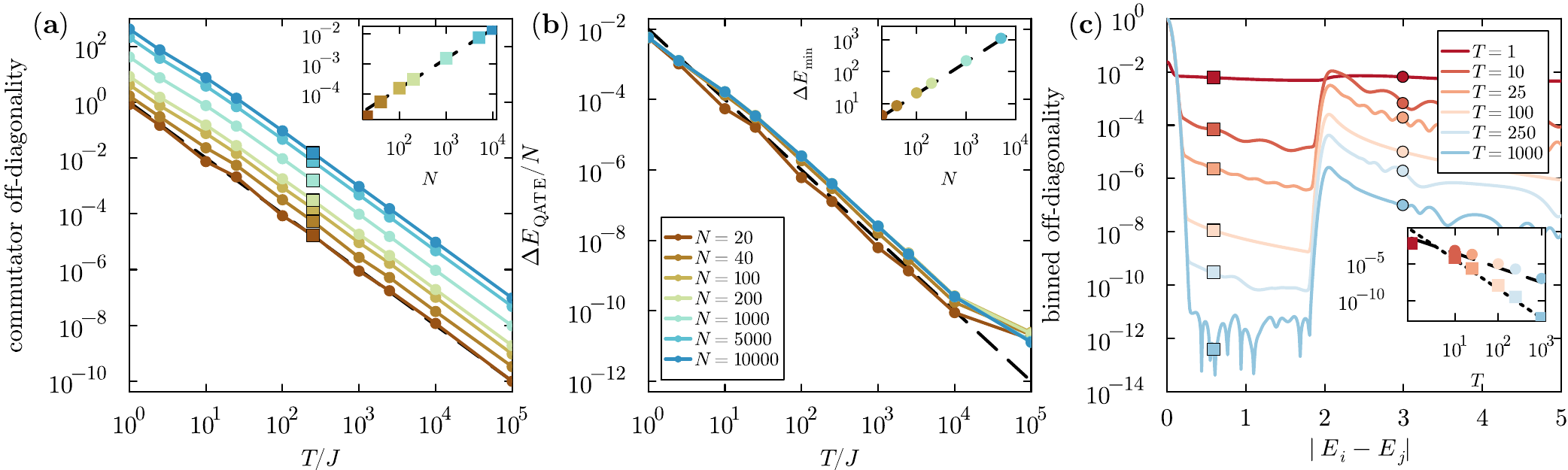}
    \caption{QATE for TFIM, evolving from $g = 1.1 \to 1.5$ at $\beta = 1$.
    (a) COD dependence on the evolution time $T$ for various system sizes $N$ along with a $1/T^2$ trendline (black dashed line). \textit{Inset}: COD scaling with system size $N$, fitted linearly (black dashed line). Square markers in the inset correspond to the square markers in the main plot. 
    (b) Dependence of the energy error normalized by the system size $N$, $\dEqate/N$, on $T$ for various $N$ along with a $1/T^2$ trendline (black dashed line). \textit{Inset}: $\dEmin$ for different system sizes, fitted linearly (black dashed line). Square markers in the inset correspond to square markers in the main plot. 
    (c) BOD dependence on the eigenstate energy difference for various $T$. Results are shown for $N = 1000$ and filter width $\delta = 0.04$. \textit{Inset}: BOD scaling with $T$, along with $1/T^2$ (black long-dash) and $1/T^4$ (black short-dash) trendlines. Square and circle markers in the inset correspond to the respective markers in the main plot. 
    }
    \label{fig:comb_tfi}
\end{figure*}

When the evolution does not cross the critical point ($g = 1$),
none of the gaps in these subsystems close, and we can analytically establish an $\mathcal{O}(N T^{-2})$ scaling for COD, $\dEqate$, and $\Delta \Var(\rhoadi)$. We verify this scaling numerically in \cref{fig:comb_tfi}, where we perform QATE from $g_{\mathrm{init}} = 1.1$ to $g_{\mathrm{final}} = 1.5$ at $\beta = 1$. 
Figures \ref{fig:comb_tfi}(a) and (b) show, respectively, COD and $\dEqate$ for various $T$ and $N$ values. Additionally, we include BOD results, obtained using filtering techniques (\cref{app:filter}), in \cref{fig:comb_tfi}(c) to examine how the off-diagonal coefficients behave for different energy eigenstates. Notably, BOD exhibits a distinct step at $\abs{E_i - E_j} = 2$. This feature is particular of the translationally invariant model studied in this section, and results from the fact that, at leading order in $T$, contributions to coherences involve the off-diagonal elements in a single $\pm k$ block (times diagonal elements in the others). Such contributions connect eigenstates separated by a $2 \epsilon_k$ energy difference, such that the minimum energy difference is determined by the magnitude of the smallest eigenmode. 
Coherences at energy differences below this threshold can only come from higher order contributions and are thus suppressed by higher powers of $T^{-1}$. This is further illustrated in \cref{app:supp_filter_res}, 
where we show that BOD contributions analytically computed up to second order successfully reproduce the observed behavior.

In the case where the evolution crosses the critical point at $g = 1$, the scaling behavior changes due to the closing gap in one of the two-level systems.  Instead of the $\mathcal{O}(N T^{-2})$ scaling observed previously, we find a Kibble-Zurek scaling of $\mathcal{O}(N T^{-1/2})$ for short times, followed by a return to the $T^{-2}$ scaling at sufficiently long times $T=\mathcal{O}(N^3)$, when the adiabatic condition is satisfied.
A more detailed discussion of this scenario can be found in \cref{app:tfi_transition}.

Even though the TFIM is particular, in that its dynamics effectively reduces to disconnected two-level systems and allows a simple analytical treatment,
 the behaviors found for the benchmarks persists for more generic systems, as we observe in the following sections.

\subsection{Isospectral Model}

To ensure the possibility of reaching the exact Gibbs state (and the associated energy), we require that $\Hi$ and $\Hf$ are isospectral, or have spectra that are exactly proportional (up to a constant shift), resulting in $\dEmin = 0$. Such models have been investigated before in \cite{Plastina2014}, which connected $\dEqate$ with the quantum relative entropy between the states $\rhoadi$ and $\rho_{\mathrm{G}}$, $\mathcal{D}(\rhoadi || \rho_{\mathrm{G}})$. Quantum relative entropy can be used to quantify the off-diagonality \cite{Plastina2019}, as well as an indistinguishability measure between the two states \cite{Audenaert2014}. From the definition of $\mathcal{D}(\rhoadi || \rho_{\mathrm{G}})$ it follows that:
\begin{align}
    \mathcal{D}(\rhoadi || \rho_{\mathrm{G}}) &= \Tr(\rhoadi (\ln(\rhoadi)-\ln(\rho_{\mathrm{G}}))) \nonumber\\
    &=\sum_k \ln(\frac{\exp(-\beta E_k^{f})}{\Zf}) \nonumber\\&\cross\left(\frac{\exp(-\beta E^i_k)}{\Zi} - \bra{E_k^f}\rhoadi \ket{E_k^f}\right)\nonumber\\
    &=\beta \dEqate,
\end{align}
where we have used the isentropic and isospectral conditions going to line two. This further indicates that $\dEqate$ is a good benchmark for investigating the quality of QATE.  

In general, the isospectral condition is not satisfied for arbitrary models. However, for free fermion models like TFIM, one can analytically compute the energy eigenmodes $\epsilon_k$ and construct another lattice Hamiltonian with identical modes. To investigate this, we keep $\Hf$ as before—$H_{\mathrm{TFI}}$ at $g = 1.5$—but engineer the initial Hamiltonian $\Hi$ as
\begin{align}
    H_{\mathrm{Z}} &= \sum_{k=1}^N \frac{\epsilon_k}{2} \sigma^z_k,
\end{align}
where $\epsilon_k$ are the eigenmodes of $\Hf$,
\begin{align}
    \epsilon_k &= 2 \sqrt{1 + g^2 + 2g\cos(2\pi k/N)}.
\end{align}
This construction ensures that $\Hi$ and $\Hf$ share the same spectrum. 
Having identical spectra guarantees that the von Neumann entropies of the Gibbs states of $\Hi$ and $\Hf$ are equal for a given $\beta$.

\begin{figure*}[t!]
    \centering
    \includegraphics[width=\linewidth]{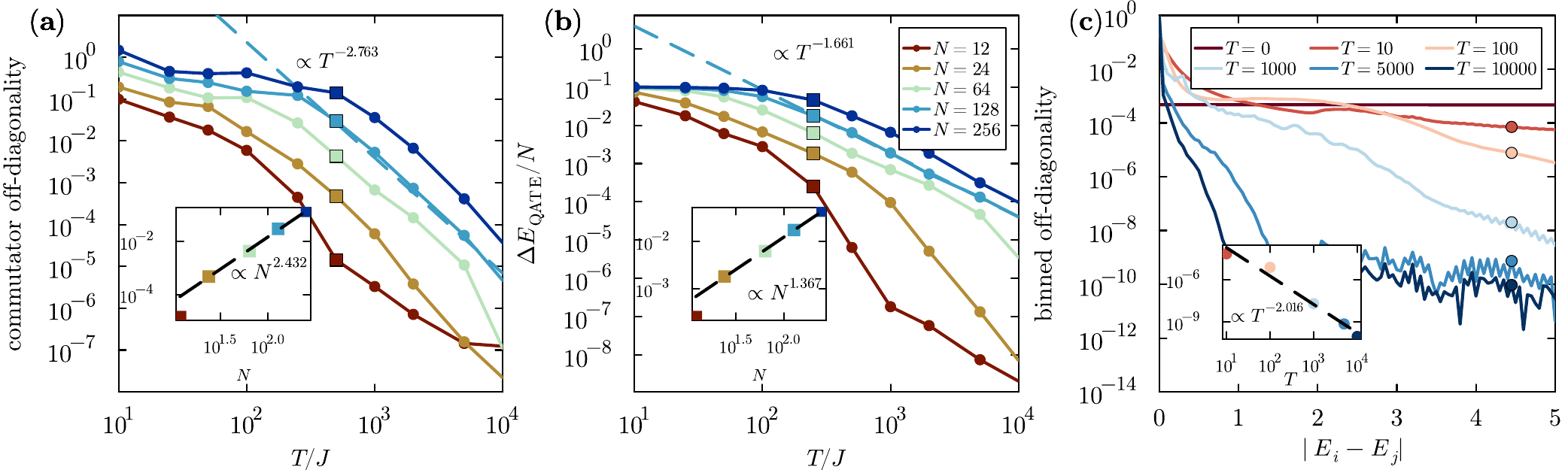}
    \caption{QATE from the isospectral $H_{\mathrm{Z}}$ to $H_{\mathrm{TFI}}$ with $J = 1$, $g = 1.5$, and $\beta = 1$.
    (a) COD dependence on the evolution time $T$ for various system sizes $N$. Blue dashed line displays a polynomial fit for $N=128$ and $T\geq250$. \textit{Inset}: system size scaling fitted at selected points ($T=500$, square markers in the inset correspond to the square markers in the main plot), with a polynomial fit (black dashed line). 
    (b) $\dEqate /N$ dependence on $T$. Blue dashed line displays a polynomial fit for $N=128$ and $T\geq250$. \textit{Inset}: system size scaling fitted at selected points ($T=250$, square markers in the inset correspond to the square markers in the main plot), with a polynomial fit (black dashed line).  
    (c) BOD dependence on the energy eigenvalue difference $\abs{E_i - E_j}$. Results are shown for $N = 256$ and filter width $\delta = 0.01$. \textit{Inset}: polynomial fit of selected points at the tails (circle markers in the inset correspond to the circle markers in the main plot). 
}
    \label{fig:isospectral_comm_e_filter}
\end{figure*}

We numerically investigate QATE from $H_{\mathrm{Z}}$ to $H_{\mathrm{TFI}}$ for various values of $T$ and $N$. 
While the model is still mapped to free fermions, the lack of translational invariance in this case limits the system sizes we can address to a few hundreds of sites. The results are shown in \cref{fig:isospectral_comm_e_filter}.
We observe that BOD tails (\cref{fig:isospectral_comm_e_filter}c) seem to behave approximately as $T^{-2}$, consistent with the previous case. However, both COD (\cref{fig:isospectral_comm_e_filter}a) and $\dEqate/N$ (\cref{fig:isospectral_comm_e_filter}b) show fairly irregular behavior for small system sizes $N$ and short times $T$. To extract the scaling, we perform a polynomial fit for data points with $T \geq 250$ and $N = 128$. This analysis for long QATE times reveals again a polynomial scaling with $N$ and $T^{-1}$. We find a COD scaling of approximately $\mathcal{O}(N^{2.4} T^{-2.8})$ and a $\dEqate$ scaling of $\mathcal{O}(N^{2.4} T^{-1.7})$. Surprisingly, the scaling with $T$ appears to be
superior to that in TFIM case, though the scaling with system size $N$ is worse. However, these results are limited to moderate system sizes, and a more systematic investigation is left for future work.

\subsection{Non-Integrable Model}\label{sec:res_non_int}

To benchmark a generic, non-integrable system, we use an Ising model with mixed fields,
\begin{equation}\label{eq:ising_mixed_fields}
    H_{\mathrm{mixed}} = \sum_{i=1}^{N-1} J \sigma^z_i \sigma^z_{i+1} + \sum_{i=1}^N h \sigma^z_i + g \sigma^x_i,
\end{equation}
where $J$, $h$, and $g$ are tunable parameters.

We initialize the system in the Gibbs state of a Hamiltonian $\Hi$ with parameters $(J, h, g)_{\mathrm{init}} = (1, 0.0, 1.05)$, which corresponds to an instance of TFIM. As the target Hamiltonian $\Hf$, we choose $(J, h, g)_{\mathrm{final}} = (1, 0.5, 1.05)$, placing the system far from integrability. In this case, QATE does not cross the zero-temperature phase transition. For this model, with $J=1, h = 0.5$ the phase transition lies at $g<1$. For a complete phase diagram of the mixed field Ising model we refer to \cite{Ovchinikov2003}.

\begin{figure*}[t!]
    \centering
    \includegraphics[width=\linewidth]{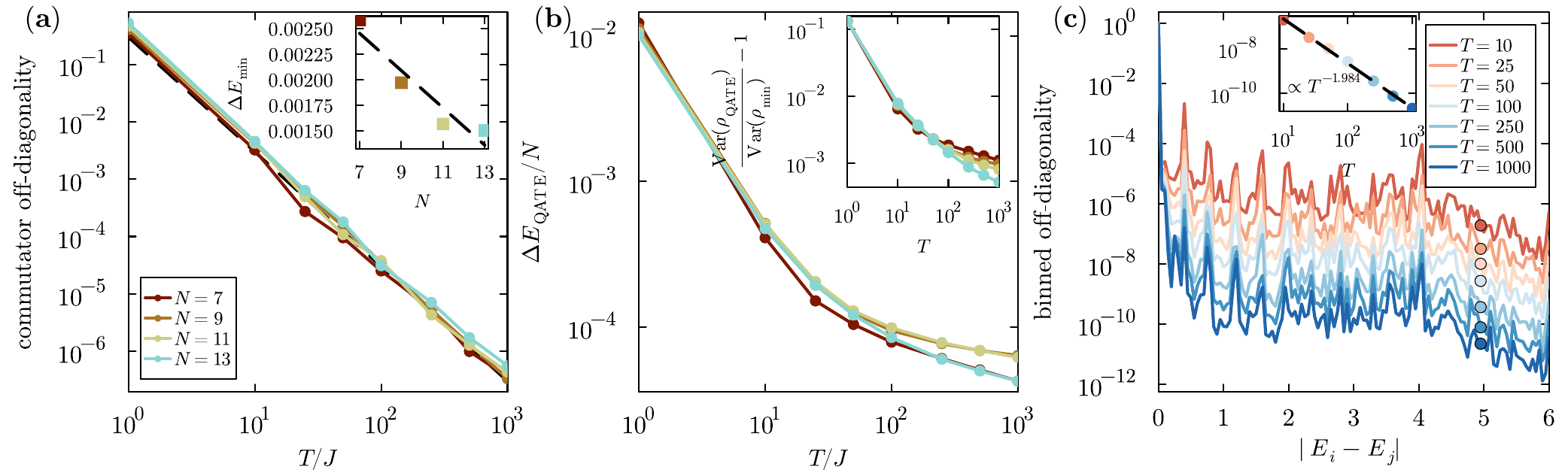}
    \caption{Ising model with mixed field evolved from $(J, h, g) = (1, 0.0, 1.05)\xrightarrow{}(1,0.5,1.05)$ at $\beta=1$. (a) Dependence of COD on the evolution time $T$ for various system sizes $N$. \textit{Inset}: $\dEmin$ for different system sizes, along with a linear fit. (b) Dependence of energy error normalized with the system size $\dEqate/N$ on $T$ for various system sizes $N$. \textit{Inset}: Dependence of the relative variance of the $\rhoadi$ with respect to $\rhomin$ on $T$. (c) BOD dependence on the energy eigenvalue difference $\abs{E_i - E_j}$ for various $T$. Inset: polynomial fit of selected points (black dashed line) at the tails (circle markers in the inset correspond to the circle markers in the main text).
    }
    \label{fig:ed_tfi_start}
\end{figure*}

In \cref{fig:ed_tfi_start}, we benchmark the performance of QATE for this model using exact diagonalization for small systems. Interestingly, both COD and BOD exhibit the familiar $T^{-2}$ scaling, consistent with the behavior observed for the integrable model in \cref{fig:comb_tfi}. Furthermore, COD (\cref{fig:ed_tfi_start}a) appears to be independent of $N$, yielding better scaling than in TFIM case. However, the scaling of $\dEqate$ with $T$ (\cref{fig:ed_tfi_start}b) flattens out at large $T$, exhibiting slower convergence compared to TFIM. This indicates that achieving energy convergence is more challenging in the non-integrable setting.

We have found that the choice of the initial state significantly affects the performance of QATE. In particular, it may be interesting to choose $\Hi$ with  $g_{\mathrm{init}} =0$, such that it consists of commuting terms and its thermal state can be easily prepared \cite{yimin_2016}. We have however found that the degenerate spectrum results in a substantially worse performance of QATE, irrespective of whether the evolution crosses or not the zero temperature phase transition.
In particular, the decrease of COD with $T$ flattens out, and the convergence of $\dEqate$ slows down even further (see \cref{app:extra_non_int} for details). 

We conclude that, for the successful application of QATE, it is important to choose an initial state that is non-degenerate. This finding is consistent with the known requirements for success in QAA. In contrast, crossing a zero-temperature phase transition does not appear to be a limiting factor, as the bulk eigenstates exhibit vanishing spectral gaps in any case.

Another reasonable question is how the protocol performs in terms of local observables. To investigate this aspect, we have computed the reduced density matrix for the three central sites of the chain and compared it to the corresponding reduced density operator for $\rhoadi$ and $\rhomin$, using the 1-norm as the distance measure.

\begin{figure}[t!]
\centering
\includegraphics[width=\linewidth]{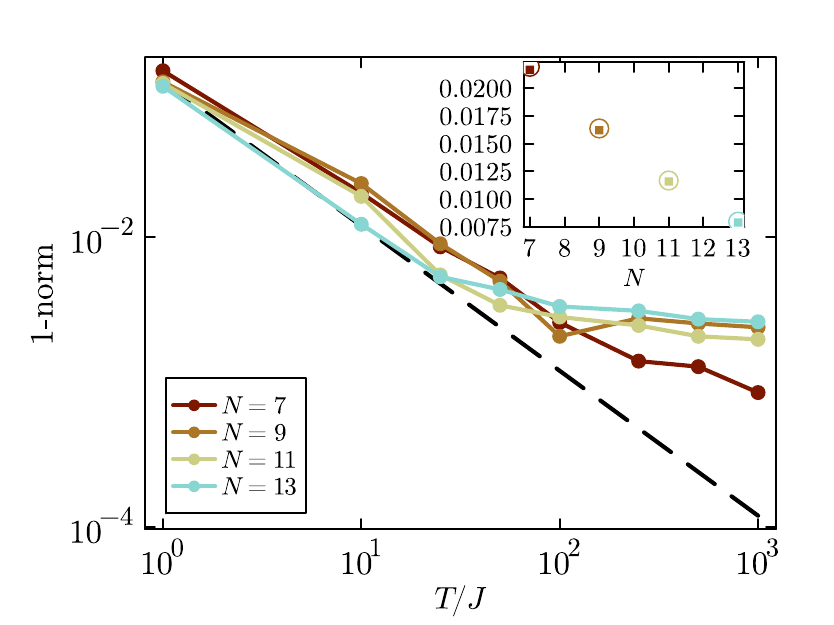}
\caption{Ising model with mixed fields evolved from $(J, h, g) = (1, 0.0, 1.05)\xrightarrow{}(1, 0.5, 1.05)$ at $\beta = 1$. Dependence of the 1-norm distance between the reduced density matrices $\rhoadi$ and $\rhomin$ on the three central sites as a function of QATE time $T$ along with a $T^{-1}$ trendline (black dashed line). The inset shows the 1-norm between $\rhomin$ and the Gibbs state at the same entropy, $\rho_{\mathrm{G}}$ (circles), and at the same energy, $\rho_{\mathrm{E}}$ (squares), as a function of system size $N$. (See \cref{fig:approach} for a visualization of $\rho_{\mathrm{G}}$ and $\rho_{\mathrm{E}}$).}
\label{fig:zzxz_obs}
\end{figure}

The results, shown in \cref{fig:zzxz_obs} display the 1-norm for different evolution time $T$ and system size $N$ values. For short times the 1-norm decays as $T^{-1}$, before flattening out. In the inset we investigate the 1-norm between the reduced matrices of the fully adiabatic state $\rhomin$ and the Gibbs states at the same entropy and energy. In both cases we observe a decrease with the system size.

\subsection{Dependence on QATE Parameters} 

The analysis in previous sections discussed the performance of QATE in different setups, but always at fixed inverse temperature, $\beta = 1$. By systematically exploring the protocol for different temperatures in the case of the isospectral model (see \cref{app:isospectral_betas}), we have found that the behavior at $\beta=1$ is in fact representative, as we observe consistent scalings over an extensive temperature range for sufficiently large systems and long times.
We choose the isospectral model for this analysis because it allows us to explore large system sizes and long evolution times and, importantly, does not exhibit the same ground-state-like behavior as QATE between two instances of TFIM. This makes the isospectral model more general for our analysis. 
As expected, 
we observe improved performance at both very large and very small $\beta$ values, corresponding to zero and maximum von Neumann entropy, respectively. Additionally, in \cref{app:temperature_non_int} we explore the temperature dependence in the non-integrable case. From \cref{fig:ed_beta_dep}, we observe that the worst-case behavior for both COD and $\dEqate$ again occurs around $\beta \approx 1$. However, in this case we find quantitatively different scaling behaviors of $\dEqate$ in the low- and high-temperature limits. In particular, at large $\beta$ we observe the expected $T^{-2}$ scaling associated with ground-state behavior, whereas at small $\beta$ the decay is slower, following an approximate $T^{-0.8}$ dependence. In both regimes, the behavior is systematic and the corresponding scaling persists even for very long QATE evolution times $T$.

For simplicity, and to facilitate easier comparison of results, we have also fixed a linear ramp $\gamma(s) = s$ in all the previous discussions.
However, in QAA, as discussed in \cref{sec:QAA}, superior error convergence can be achieved by using a smooth evolution schedule $\gamma(s)$ with vanishing derivatives at the start and end points. 
A similar improvement is possible in QATE by modifying the ramp.
In \cref{app:smooth_ramp}, we examine a specific schedule where the first three derivatives vanish at $s = 0$ and $s = 1$. For QATE applied to TFIM without a phase transition crossing, we observe significantly improved scaling of COD and $\dEqate$ with $T$, improving from $\mathcal{O}(T^{-2})$ to approximately $\mathcal{O}(T^{-8})$ in the large-$T$ regime. These results suggest that employing a smooth evolution schedule can lead to further improvements in QATE performance, enabling better results at shorter evolution times $T$.
\section{Conclusions}\label{sec:conclusions}

In this work, we investigated the quasi-adiabatic protocol (QATE) as a method for probing thermal properties of quantum systems at finite temperatures. Building on insights from QAA, we proposed applying a finite-time adiabatic evolution starting from an initial Gibbs state. We identify the diagonality of the state, the difference in average energy with respect to the adiabatic limit
$\dEqate$, and the energy variance as important benchmarks for success.

We analytically demonstrate the convergence of these quantities in the special case of TFIM, where the dynamics reduce to a collection of two-level systems. Furthermore, for both integrable and non-integrable models where such a separation is not possible, we perform numerical simulations and nevertheless observe behavior similar to that of TFIM. In particular, the off-diagonality measures at the end of QATE---COD and BOD---exhibit a characteristic $T^{-2}$ scaling for both integrable and non-integrable translationally invariant models, with the energy error $\dEqate$ and variance converging more slowly in the non-integrable cases.

We conclude that this behavior persists across temperatures, with improved performance in the zero- and infinite-temperature limits. Our study highlights that the choice of initial state is critical: degenerate initial states hinder the success of the QATE, while crossing a zero-temperature phase transition does not strongly affect performance due to the presence of vanishing spectral gaps in the bulk. Future work could aim to extend these findings to larger non-integrable systems and obtain analytical guarantees for COD and BOD quantities in the generic case. The simplicity of the QATE makes it a promising near-term approach to explore across quantum simulation platforms.

\acknowledgments{We thank Sirui Lu, Benjamin F. Schiffer, Yilun Yang and Samuel J. Garratt for helpful discussions. We
acknowledge the support from the German Federal Ministry of Education and Research (BMBF) through FermiQP (Grant No. 13N15890) within the funding program quantum technologies—from basic research to market. This research is
part of the Munich Quantum Valley (MQV), which is supported by the Bavarian state government with funds from the Hightech Agenda Bayern Plus.
This work was partially supported by the Deutsche Forschungsgemeinschaft (DFG, German Research Foundation) under Germany's Excellence Strategy -- EXC-2111 -- 390814868;  
and by the EU-QUANTERA project TNiSQ (BA 6059/1-1).}
\bibliographystyle{quantum}
\bibliography{main}

\appendix
\section{Exact solution of the TFI model}\label{app:tfi-exact}

Consider the transverse field Ising model on $N$ sites ($N$ even) with periodic boundary conditions,
\begin{equation}
    H_{\mathrm{TFI}} = -\sum_{n=1}^{N-1}\sigma_n^x\sigma_{n+1}^x - g\sum_{n=1}^{N}\sigma_n^z - J_{bc}\sigma_N^x\sigma^x_1.
\end{equation}
Applying the Jordan-Wigner transformation \cite{JordanWigner}, the Hamiltonian becomes
\begin{align}
\label{eq:TFI_fermions}
\hat{H} &= -\sum_{n=1}^{N-1}(a_n^\dagger - a_n)(a_{n+1} + a_{n+1}^\dagger) \nonumber\\
&\quad + P J_{bc} (a_N^\dagger - a_N)(a_1 + a_1^\dagger) - g \sum_{n=1}^{N}(a_n^\dagger a_n - a_n a_n^\dagger),
\end{align}
where $P = \prod_{n=1}^{N}(1-2a_n^\dagger a_n)$ is the total parity operator.

To map the TFI model to a fermionic quadratic Hamiltonian, we must consider the parity sectors separately.  
Throughout this work, we choose an even system size $N$ and impose periodic boundary conditions in \cref{eq:TFI_fermions}, setting $P J_{bc} = -1$.  
This results in $J_{bc} = 1$ ($-1$) for the odd (even) occupation sectors of \cref{eq:TFI}. Applying a Fourier transform, the Hamiltonian becomes
\begin{align}
\label{eq:TFI_fourier}
H &= -2\sum_{k > 0} (g+\cos(2\pi k/N))(f_k^\dagger f_k - f_k f_k^\dagger) \nonumber\\
&\quad + i \sin(2\pi k/N)(f_k^\dagger f_{-k}^\dagger - f_{-k} f_k) \nonumber\\
&\quad - (1+g)(f_0^\dagger f_0 - f_0 f_0^\dagger) \nonumber\\
&\quad - (g-1)(f_{-N/2}^\dagger f_{-N/2} - f_{-N/2} f_{-N/2}^\dagger),
\end{align}
where
\begin{equation}
	f_k = \frac{1}{\sqrt{N}}\sum_{j=1}^{N} e^{i\frac{2\pi}{N}kj} a_j, \quad
	f_k^\dagger = \frac{1}{\sqrt{N}}\sum_{j=1}^{N} e^{-i\frac{2\pi}{N}kj} a_j^\dagger,
\end{equation}
and $k = 1, 2, \dots, N/2 - 1$. The Hamiltonian couples operators with momenta $k$ and $-k$, making it block-diagonal but not fully diagonal.  
To fully diagonalize the model, we diagonalize each $k$-block separately following \cite{Surace_2022}.
Each block
\begin{equation}
h_k = \begin{pmatrix} -2(g+\cos(2\pi k/N)) & 2i \sin(2\pi k/N) \\ -2i \sin(2\pi k/N) & 2(g+\cos(2\pi k/N)) \end{pmatrix}
\end{equation}
is diagonalized by the Bogoliubov transformation \cite{bogoljubov1958new}
\begin{equation}
\label{eq:Bogoliubov_transform}
U_k = \begin{pmatrix} s_k & -i t_k \\ -i t_k & s_k \end{pmatrix},
\end{equation}
where
\begin{align}
s_k &= \frac{\sin(2\pi k/N)}{ \sqrt{ \epsilon_k(\epsilon_k/2 + g + \cos(2\pi k/N)) } }, \\
t_k &= \frac{ \epsilon_k/2 + g + \cos(2\pi k/N) }{ \sqrt{ \epsilon_k(\epsilon_k/2 + g + \cos(2\pi k/N)) } },
\end{align}
and
\begin{equation}
\label{eq:tfi_eigenmodes}
\epsilon_k = 2 \sqrt{1 + g^2 + 2g\cos(2\pi k/N)}.
\end{equation}
The transformation brings each $h_k$ into diagonal form
\begin{equation}
U_k^\dagger h_k U_k = \begin{pmatrix} \epsilon_k & 0 \\ 0 & -\epsilon_k \end{pmatrix}.
\end{equation}
The Fourier modes $f_k$ transform to the new fermionic operators $b_k$ via
\begin{align}
f_k &= s_k b_k - i t_k b_{-k}^\dagger, \\
f_{-k}^\dagger &= s_k b_{-k}^\dagger - i t_k b_k.
\end{align}
Substituting these relations into the full Hamiltonian gives
\begin{align}
\label{diagonal_form}
H &= \sum_{k\neq -N/2, 0} \frac{\epsilon_k}{2}(b_k^\dagger b_k - b_k b_k^\dagger) \nonumber\\
&\quad - (1-g)(f_0^\dagger f_0 - f_0 f_0^\dagger) \nonumber\\
&\quad - (g-1)(f_{-N/2}^\dagger f_{-N/2} - f_{-N/2} f_{-N/2}^\dagger),
\end{align}
where the allowed momenta are $k = -N/2, \dots, N/2-1$ for our choice of boundary conditions.

\subsection{Solving a single mode system}\label{sec:single_mode}

First, note that the modes $f_0$ and $f_{-N/2}$ remain unchanged under variation of the coupling $g$, since they do not transform under the Bogoliubov transformation. Consequently, these blocks remain diagonal under the QATE, provided that the initial state in the block is diagonal. We work in the basis $(\ket{0}_f, f_{0/(-N/2)}^{\dagger}\ket{0}_f)$ for these modes. For the remaining $k$ values, the dynamics are confined to blocks that couple momenta $\pm k$. It is thus natural to investigate each block separately. We work in the basis
\begin{equation}
\vec{b} = (\ket{0}_b, b_{-k}^\dagger \ket{0}_b, b_{k}^\dagger \ket{0}_b, b_{k}^\dagger b_{-k}^\dagger \ket{0}_b)^T.
\end{equation}
In this basis, the Hamiltonian for the sector reads
\begin{equation}
\label{eq:diagonal_hk}
D_k^b = \text{diag}(-\epsilon_k, 0, 0, \epsilon_k),
\end{equation}
and the thermal state is given by
\begin{align}
\rho_k^b &= \frac{1}{\mathcal{Z}_k} \text{diag}(e^{\beta \epsilon_k}, 1, 1, e^{-\beta \epsilon_k}), \\
\mathcal{Z}_k &= \left(e^{-\beta \epsilon_k/2} + e^{\beta \epsilon_k/2}\right)^2.
\end{align}
This defines the initial state for the QATE protocol. We aim to probe the effect of time evolution under a Hamiltonian with a different coupling $g'$ (a quench) on the initial thermal state. This procedure serves as a building block for performing the full QATE (in its Trotterized form). We introduce the Fourier basis
\begin{equation}
\vec{f} = (\ket{0}_f, f_{-k}^\dagger \ket{0}_f, f_{k}^\dagger \ket{0}_f, f_{k}^\dagger f_{-k}^\dagger \ket{0}_f)^T,
\end{equation}
which will serve as a reference when transforming between different couplings. Here, $\ket{0}_f$ denotes the Fourier vacuum annihilated by $f_k$ and $f_{-k}$, as opposed to the Bogoliubov vacuum $\ket{0}_b$ annihilated by $b_k$ and $b_{-k}$.

Using \cref{eq:Bogoliubov_transform}, we establish the transformation relationships between the $\vec{b}$ and $\vec{f}$ bases. The vacua are related via
\begin{align}
b_{\pm k}\ket{0}_b = (s_k f_{\pm k} \pm i t_k f_{\mp k}^\dagger) \ket{0}_b = 0,
\end{align}
which is solved by
\begin{equation}
\ket{0}_b = s_k \ket{0}_f - i t_k f_k^{\dagger} f_{-k}^{\dagger} \ket{0}_f.
\end{equation}
Applying the relevant operators to the vacuum yields
\begin{equation}
\label{eq:Tk}
\vec{b} = \begin{pmatrix}
s_k & 0 & 0 & -i t_k \\
0 & 1 & 0 & 0 \\
0 & 0 & 1 & 0 \\
-i t_k & 0 & 0 & s_k
\end{pmatrix} \vec{f} = T_k \vec{f}.
\end{equation}
Note that in this analysis we have omitted explicitly writing the dependence of $\vec{b}$ on the coupling $g$. Going forward, we will indicate this dependence when necessary.

\subsection{Performing a quench}

We consider time-evolving a Gibbs state prepared at coupling value $g$ under a Hamiltonian with coupling value $g'$. We denote the respective bases as $\vec{b}$ and $\vec{b'}$. In each case, the transformation to the Fourier basis is given by
\begin{align*}
    \vec{b} &= T_k \vec{f},\\
    \vec{b'} &= T_k' \vec{f},
\end{align*}
which allows us to directly transform between the two bases. Thus, the time-evolved state can be written as
\begin{align}
    &\exp(-iH(g')t) \rho_k(g) \exp(iH(g')t) \\
    &= T_k' \exp(-iD_k(g')t) T_k'^{\dagger} T_k \rho_k^b(g) T_k^{\dagger} T_k' \exp(iD_k(g')t) T_k'^{\dagger}\nonumber,
\end{align}
giving the final result expressed in the Fourier basis. The QATE in its Trotterized form can then be constructed as a sequence of such quenches, each with varied coupling values $g'$.

\subsection{Performing Trotterized QATE of TFI}
\label{sec:QATE_free_fermions}

We consider a Trotterized version of the QATE with $M$ steps and a time step $\tau$, interpolating from $g_0$ to $g_1$. The instantaneous coupling is given by
\begin{equation}
    g(s) = (1-\gamma(s))g_0 + \gamma(s)g_1,
\end{equation}
where $s = j/M$ at step $j$.

In the Fourier basis, the evolution reads
\begin{align}
    \rho_k^1 &= \mathcal{U}_{M,\tau} \rho_k^0 \mathcal{U}_{M,\tau}^{\dagger}, \quad \text{where} \\
    \mathcal{U}_{M,\tau} &= \prod_{j=1}^{M} \exp(-i h_k(s) \tau) \nonumber\\&= \prod_{j=1}^{M} T_k(s) \exp(-i D_k(s) \tau) T_k^{\dagger}(s), \\
    \rho_k^0 &= T_k(0) \rho_k^b(g_0) T_k^{\dagger}(0),
\end{align}
which defines the Trotterized QATE for this model.

\subsection{Off-diagonality magnitude}

To establish the scaling of the off-diagonal elements with the total adiabatic evolution time $T$, it is sufficient to work without explicitly performing the Trotterization. First, note that $T_k$ in \cref{eq:Tk} is block diagonal, with the middle block being the identity and the outer block corresponding to $U_k$. Consequently, the inner modes never mix and thus do not contribute to off-diagonality. The evolution of each $\pm k$ block therefore separates into two independent two-level systems. The middle block remains proportional to the identity throughout, while the outer block can be treated as a two-level system with spectral gap $2\epsilon_k$ (see \cref{eq:diagonal_hk}).

Thus, the evolved density matrix takes the form
\begin{equation}
\label{eq:rho_final_form}
    \rho_k(1) =
    \begin{pmatrix}
        a_f & 0 & 0 & c_k \\
        0 & \calZ_k^{-1} & 0 & 0 \\
        0 & 0 & \calZ_k^{-1} & 0 \\
        c_k^* & 0 & 0 & 1-2\calZ_k^{-1}-a_f
    \end{pmatrix},
\end{equation}
where the only off-diagonal elements are $c_k$. To bound $c_k$ we can directly apply the results from \cref{sec:QAA} and \cref{eq:qaa} to conclude that,
\begin{equation}
    \label{eq:ck_scaling}
    |c_k|^2 = \mathcal{O}\left( \max_s \frac{|\partial_s \epsilon_k(g(s))|^4}{\epsilon_k(g(s))^6 T^2} \right) = \mathcal{O}\left( \frac{1}{T^2} \right),
\end{equation}
provided that the evolution path does not cross the critical point at $g=1$.

\subsection{Calculating the COD}

The resulting matrix in the final eigenbasis is given by
\begin{equation}
    \rho^{b}(1) = \rho_{-N/2}^b \otimes \bigotimes_{k=0}^{N/2-1} \rho_k^b,
\end{equation}
where the Hamiltonian in the same basis is
\begin{equation}
    D^b = \sum_k D_k^b,
\end{equation}
with $D_k^b$ acting only on the subsystem of the $\pm k$ modes and as the identity on the rest. The commutator off-diagonality (COD) can be calculated as
\begin{align}
    \label{eq:cod_tfi}
    &\Tr\left( \comm{\rho}{H}^2 \right) = \sum_k \Tr\left( \comm{\rho_k^b}{D_k^b}^2 \otimes_{j \neq k} \rho_j^2 \right) \\
    &= \sum_k \Tr\left( \comm{\rho_k^b}{D_k^b}^2 \right) \prod_{j \neq k} \Tr(\rho_j^2) \nonumber\\
    &= \Tr(\rho^2) \sum_k \frac{ \Tr\left( \comm{\rho_k^b}{D_k^b}^2 \right) }{ \Tr(\rho_k^2) } \nonumber\\
    &= - \Tr(\rho^2) \sum_{k=1}^{N/2-1} \epsilon_k^2 \frac{8 |c_k|^2}{a_k^2 + (1-2\calZ_k^{-1} - a_k)^2 + 2 |c_k|^2},\nonumber
\end{align}
where we have used the form of $\rho_k^b$ from \cref{eq:rho_final_form}.

Using the scaling of $|c_k|^2$ from \cref{eq:ck_scaling} and the fixed purity dependence from \cref{eq:fixed_N_dep}, we can establish that the COD is bounded as
\begin{equation}
\label{eq:comm_scaling}
    -\frac{\Tr\left( \comm{\rho}{H}^2 \right)}{\Tr(\rho^2)} = \mathcal{O}\left( \frac{N}{T^2} \right),
\end{equation}
assuming a linear schedule $\gamma(s) = s$ and avoiding the transition at $g=1$.

\subsection{Energy, off-diagonality, and variance}
\label{app:tfi_variance}
In this section we show that $\dEqate$, off-diagonality, and the energy variance are connected in the TFI. To illustrate this we consider a single mode in the basis of the Hamiltonian $D_k = \diag(-\epsilon_k, \epsilon_k)$, where the constant middle block is omitted for simplicity. Consider the diagonal mode as

\begin{equation}\rhomin^k=
\begin{pmatrix}
    a_k & 0 \\
    0 & d_k-a_k
\end{pmatrix},
\end{equation}
and the one obtained under QATE with some off-diagonality as:
\begin{equation}\rhoadi^k=
\begin{pmatrix}
    b_k & c_k \\
    c_k^* & d_k-b_k
\end{pmatrix},
\end{equation}
where $d_k = 1-2\calZ_k^{-1}$ (\cref{eq:rho_final_form}).
The relevant quantities to consider are:
\begin{align}
    E^k_{\min} &= \epsilon_k(d_k-2a_k),\\ \Var^k(\rhomin) &= [d_k - (d_k-2a_k)^2]\epsilon^2_k,\\ 
    E^k_{\qate} &= \epsilon_k(d_k-2b_k),\\
    \Var^k(\rhoadi) &= [d_k - (d_k-2b_k)^2]\epsilon^2_k,
\end{align}
and their respective differences being given by:
\begin{align}
    \dEqate^k&=2\epsilon_k(a_k-b_k)\\
    \Delta \Var^k_{\qate} &= \epsilon_k^2[(d_k-2b_k)^2-(d_k-2a_k)^2],\nonumber\\
    &=2\dEqate^k(E^k_{\min}+E^k_{\qate})
\end{align}
To see how these difference depend on the off-diagonality $c_k$, we use purity $\Tr(\rho_k^2)$, which is a conserved quantity throughout the QATE:
\begin{align}
    &a_k^2 +(d_k-a_k)^2 = b_k^2 +(d_k-b_k)^2 +2\abs{c_k}^2,\nonumber\\
    &(a_k-b_k)(2d_k-a_k-b_k)=\abs{c_k}^2,\nonumber\\
    &\dEqate^k (2d_k\epsilon_k + E^k_{\min}+E^k_{\qate})=4\abs{c_k}^2\epsilon_k^2.
\end{align}
This leads to
\begin{align}\label{eq:emin_tfi}
    E^k_{\qate}{}^2-E^k_{\min}{}^2 &= 4\abs{c_k}^2\epsilon_k^2-4d_k \epsilon_k \dEqate^k,\nonumber\\
    2E^k_{\min}\dEqate^k &\approx 4\abs{c_k}^2\epsilon_k^2-4d_k \epsilon_k \dEqate^k,\nonumber\\
    \dEqate^k &= \frac{2\abs{c_k}^2\epsilon_k^2}{E^k_{\min}+2d_k\epsilon_k}
\end{align}
in the limit of small $\dEqate^k$, and similarly
\begin{align}
    \label{eq:var_tfi}
    \Delta \Var^k_{\qate} = 8\abs{c_k}^2\epsilon_k^2-4d_k \epsilon_k \dEqate^k.
\end{align}
Thus, we have shown that both $\dEqate^k$ and $\Delta \Var^k_{\qate}$, scale as $\abs{c_k}^2$, implying the same scaling with $T$ as in the COD \cref{eq:cod_tfi}. Furthermore, for the system size dependence it is easy to see that $\dEqate = \sum \dEqate^k$, as well as that the variance of each block is additive since:

\begin{align}
    &\Tr(\rho H^2) - \Tr(\rho H)^2 \nonumber\\
    &=\sum_k \Tr(D_k^2 \rho_k) + \sum_{k, j, \, k \neq j} \Tr(D_k \rho_k) \Tr(D_j \rho_j)\nonumber\\
    &-\sum_k\Tr(D_k\rho_k) \sum_j\Tr(D_j\rho_j)\nonumber\\
    &= \sum_k \left[ \Tr(D_k^2 \rho_k) - \Tr(D_k \rho_k)^2 \right] = \mathcal{O}(N).
\end{align}
This establishes that in the case of TFIM, COD, $\dEqate$, and $\Delta \Var_{\qate}$ are $\poly(N,T^{-1})$ and all scale with \cref{eq:comm_scaling}.

\section{Results from a Gaussian approximation}\label{app:gaussian}

In this appendix, we analyze the properties of $\rhomin$ using the approximation that the density of states of a generic local Hamiltonian approaches a Gaussian form in the thermodynamic limit \cite{hartmann2004gaussian, anshu2016concentration, kuwahara2016}.  
We assume a Gaussian density of states, a traceless Hamiltonian and $\beta\geq0$. In this case
\begin{equation}
    g(E) = \frac{\mathcal{D}}{\sqrt{2\pi \sigma^2}} \exp\left(-\frac{E^2}{2\sigma^2}\right),
\end{equation}
where $\mathcal{D} = 2^N$ is the dimensionality of the system. The corresponding Hamiltonian is given by:
\begin{equation}
    H = \int_{-\infty}^{\infty} g(E) E \ket{E}\bra{E} \, dE.
\end{equation}
We are interested in the Gibbs state:
\begin{equation}
    \rho = \mathcal{Z}^{-1} \int_{-\infty}^{\infty} g(E) e^{-\beta E} \ket{E}\bra{E} \, dE,
\end{equation}
with the partition function given by:
\begin{align}
    \mathcal{Z} &= \int_{-\infty}^{\infty} e^{-\beta E} g(E) \, dE =\mathcal{D} \exp\left(\frac{\beta^2 \sigma^2}{2}\right).
\end{align}
The energy is:
\begin{align}
    \Tr(\rho H) &= \mathcal{Z}^{-1} \int_{-\infty}^{\infty} E e^{-\beta E} g(E) \, dE = -\beta \sigma^2,
\end{align}
and the variance can be found from:
\begin{align}
    \Tr(\rho H^2) &= \mathcal{Z}^{-1} \int_{-\infty}^{\infty} E^2 e^{-\beta E} g(E) \, dE = \sigma^2 + \Tr(\rho H)^2,
\end{align}
yielding variance of $\sigma^2$, as expected for a Gaussian distribution. We are interested in comparing the initial and final Hamiltonians, $\Hi$ and $\Hf$, with variances $\sigmai$ and $\sigmaf$, respectively. In particular, we want to evaluate $\Em$ and $\Eg$, as well as the variance.
\begin{align}
    \Em = \Zi^{-1} \int_{-\infty}^{\infty} E_{\mathrm{final}}(E) e^{-\beta E} g_{\mathrm{init}}(E) \, dE.
\end{align}
We assume an ideal case where each eigenstate of $\Hi$ gets mapped onto a single one of $\Hf$, without any level crossings in the path, such that the order of eigenstates is preserved. Then we can find the eigenstate correspondence by equating the cumulative probabilities:
\begin{align}
    \int_{-\infty}^{E_{\mathrm{init}}} g_{\mathrm{init}}(E) \, dE &= \int_{-\infty}^{E_{\mathrm{final}}} g_{\mathrm{final}}(E) \, dE,\nonumber \\
    \int_{-\infty}^{E_{\mathrm{init}}/\sqrt{2\sigmai^2}} e^{-x^2} \, dx &= \int_{-\infty}^{E_{\mathrm{final}}/\sqrt{2\sigmaf^2}} e^{-x^2} \, dx,
\end{align}
which leads to
\begin{equation}
    E_{\mathrm{final}} = \frac{\sigmaf}{\sigmai} E_{\mathrm{init}}.
\end{equation}
This allows us to compute $\Em$:
\begin{align}
    \Em = \frac{\sigmaf}{\sigmai} \Tr(\rhoi \Hi) = -\beta \sigmai \sigmaf.
\end{align}
To obtain the energy of a Gibbs state of $\Hf$ at the same von Neumann entropy $S$, we first determine how the entropy scales with the inverse temperature $\beta$:
\begin{align}
    S[\beta] &= -\Tr(\rho \ln(\rho))\nonumber\\&= -\mathcal{Z}^{-1} \int_{-\infty}^{\infty} e^{-\beta E} g(E) [-\beta E - \ln(\mathcal{Z})] \, dE\nonumber\\
    &=-\mathcal{Z}^{-1} \int_{-\infty}^{\infty} e^{-\beta E} g(E) [-\beta E - \frac{\beta^2 \sigma^2}{2}-\ln(\mathcal{D})] \, dE
    \nonumber\\&= -\left[\beta^2 \sigma^2 - \frac{\beta^2 \sigma^2}{2} - \ln(\mathcal{D})\right]\nonumber\\
    &=\ln{\mathcal{D}} - \frac{\beta^2\sigma^2}{2},
\end{align}
which implies that
\begin{equation}
    \beta_{\mathrm{final}} = \frac{\sigmai}{\sigmaf} \beta_{\mathrm{init}},
\end{equation}
to satisfy the isentropic condition. Using this relation, we find:
\begin{align}
    \Eg &= \Tr(\rho_{\mathrm{final}}^{(\mathrm{iso})} \Hf) \\
                   &= -\beta_{\mathrm{final}} \sigmaf^2 = -\beta_{\mathrm{init}} \sigmai \sigmaf, \nonumber\\
    \dEmin &= \Em - \Eg = 0,
\end{align}
demonstrating that $\dEmin$ vanishes in this limit for generic local Hamiltonians.
To evaluate the variance of $\rhomin$, we again use the same results:
\begin{align}
    \Tr(\rhomin \Hf^2) = \frac{\sigmaf^2}{\sigmai^2} \Tr(\rho_{\mathrm{init}} \Hi^2) = \sigmaf^2 + \Em^2,
\end{align}
which yields a variance of $\sigmaf^2$, equal to that of a Gibbs state of the target system $\Hf$.

Using the result that the density of states of a local Hamiltonian approaches a Gaussian in the thermodynamic limit, we have shown that $\dEmin$ vanishes, and that the fully adiabatic state $\rhomin$ exhibits the same energy variance as a Gibbs state of the target system.

\section{Filter Derivation}
\label{app:filter}

In this section, we derive how to design a filter to probe the off-diagonality of a state $\rho$ in the eigenbasis of $H$. We proceed as follows:
\begin{equation}
    \rho = \sum_{ij} c_{ij} \ket{E_i}\bra{E_j}.
\end{equation}
We are interested in probing the function
\begin{equation}
    f(\omega) = \sum_{ij} \abs{c_{ij}}^2 \delta\big((E_i - E_j) - \omega\big).
\end{equation}
This can be expanded as
\begin{align}
    f(\omega) &= \sum_{ij} c_{ij} c_{ij}^* \delta\big((E_i - E_j) - \omega\big) \nonumber\\
    &= \sum_{ij} c_{ij} c_{ji} \delta\big((E_i - E_j) - \omega\big).
\end{align}
We use the identity
\begin{equation}
    \int_{-\infty}^{\infty} \! dt \, e^{i\omega t} = 2\pi \delta(\omega),
\end{equation}
and note that
\begin{align}
    &\exp(-iHt)\rho \exp(iHt) e^{i\omega t} \nonumber\\
    &= \sum_{ij} e^{-it(E_i-E_j-\omega)} c_{ij} \ket{E_i}\bra{E_j}.
\end{align}
Taking the trace of this expression gives
\begin{align}
    &\Tr\left( \exp(-iHt) \rho \exp(iHt) \rho \right) e^{i\omega t} \nonumber\\
    &= \sum_{ijk} e^{-it(E_i - E_j - \omega)} \Tr\left(c_{ij} c_{jk} \ket{E_i}\bra{E_k}\right)\nonumber \\
    &= \sum_{ij} e^{-it(E_i - E_j - \omega)} c_{ij} c_{ji} \nonumber\\
    &= \sum_{ij} e^{-it(E_i - E_j - \omega)} \abs{c_{ij}}^2,
\end{align}
and integrating over time yields
\begin{align}
    &\int_{-\infty}^{\infty} \! dt \, \Tr\left( \exp(-iHt) \rho \exp(iHt) \rho \right) e^{i\omega t} \nonumber\\
    &= \sum_{ij} \abs{c_{ij}}^2 \int_{-\infty}^{\infty} \! dt \, e^{-it(E_i - E_j - \omega)}\nonumber \\
    &= \sum_{ij} \abs{c_{ij}}^2 \delta(E_i - E_j - \omega) = f(\omega),
\end{align}
which is exactly the quantity we are seeking.

However, to obtain the exact delta function, infinite time evolution would be required, which is not practically feasible. Instead, we use the filtering algorithm from \cite{Lu_2021}, which shows how to approximate the delta function by a narrow Gaussian over a finite domain. Here we briefly review the relevant parameters for the filtering.

The filter is given by
\begin{align}
    P_{\delta}(E) = \exp\left(-\frac{X^2}{2\delta^2}\right) \approx \sum_{m = -R}^{R} c_m \exp(-iXt_m),
\end{align}
where the approximation error is bounded by $2\exp(-x^2/2)$. The parameters are defined as:
\begin{itemize}
    \item $R = x \mathcal{N} / \delta$,
    \item $t_m = 2m / \mathcal{N}$,
    \item $c_m = \frac{1}{2^\mathcal{M}} \binom{\mathcal{M}}{\mathcal{M}/2 - m }$,
\end{itemize}
where $\mathcal{M} = (R/x)^2$ must be even. This approximation is valid within the domain $[-\mathcal{N}\pi/2, \mathcal{N}\pi/2]$. Since here we aim to approximate a very sharp signal, we can choose $\mathcal{N}$ to be independent of the system size. In practice, if one performs time evolution from $-\Tf$ to $\Tf$ in increments of $dt$, then
\begin{align}
    R &= \Tf / dt, \\
    \mathcal{N} &= 2 / dt, \\
    \delta &= 2x / \Tf.
\end{align}
Thus, the maximum evolution time $\Tf$ controls the width of the filter, while the time evolution graining $dt$ determines the available exploration domain, as expected from basic signal processing considerations.

\section{Supplementary results}\label{app:supp_res}

\subsection{Interpretation of the TFIM BOD Structure}
\label{app:supp_filter_res}

\begin{figure}
    \centering
    \includegraphics[width=\linewidth]{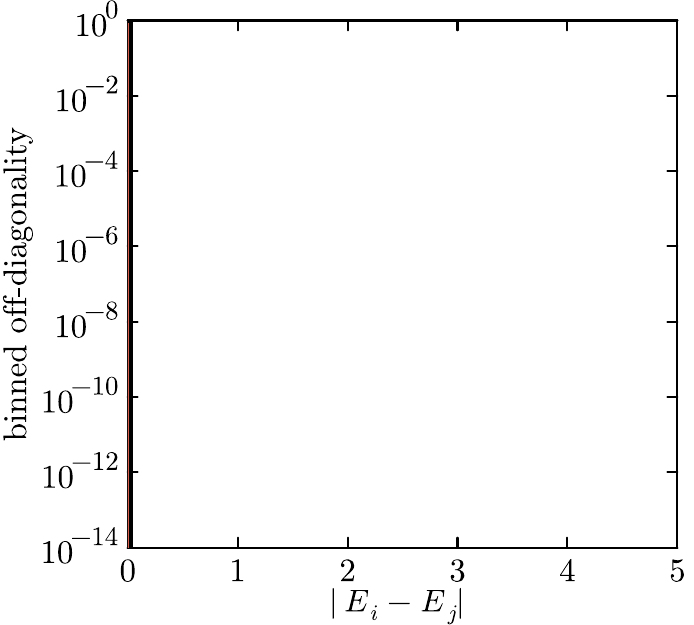}
    \caption{QATE for the TFIM, evolving from $g = 1.1 \to 1.5$ at $\beta = 1$. BOD as a function of the eigenstate energy difference $\abs{E_i - E_j}$ for $T = 100$ and $N = 1000$. The BOD is calculated up to first-order ($T^{-2}$) and second-order ($T^{-4}$) contributions and displayed as a histogram. We use bins of width $2\delta = 0.08$, consistent with the results shown in \cref{fig:comb_tfi}~(c). Additionally, the pink solid line shows the BOD obtained from filtering for the same parameters ($N = 1000$, $T = 100$, and $\delta = 0.04$) as in \cref{fig:comb_tfi}~(c).}
    \label{fig:supp_tfi_filter}
\end{figure}

In this appendix, we calculate the BOD exactly and display it in \cref{fig:supp_tfi_filter} to explain the step structure observed in \cref{fig:comb_tfi}~(c) at $|E_i - E_j| = 2$. To compute the off-diagonality exactly, we exploit the fact that the QATE for the TFIM reduces to a collection of two-level problems (see \cref{app:tfi-exact}). The BOD can then be assessed by calculating the contributions from each of the two-level blocks and combining them to a desired accuracy.

At first order ($T^{-2}$), the off-diagonality is limited to contributions within each block. This leads to eigenstate energy differences of $2\epsilon_k$, with the minimum value $\min_k(2\epsilon_k) = 4(g - 1)$. For the case considered, $g = 1.5$, this yields the observed step at $|E_i - E_j| = 2$. 

At second order ($T^{-4}$), off-diagonality arises from combinations involving pairs of distinct blocks. This leads to energy differences of the form $|E_i - E_j| = |\epsilon_k \pm \epsilon_l|$ for all $k \neq l$, allowing the second-order contributions to take a broader range of values.

\subsection{TFIM traversing \texorpdfstring{$g=1$}{g = 1}}
\label{app:tfi_transition}

\begin{figure*}[t]
\centering
\includegraphics[width=\linewidth]{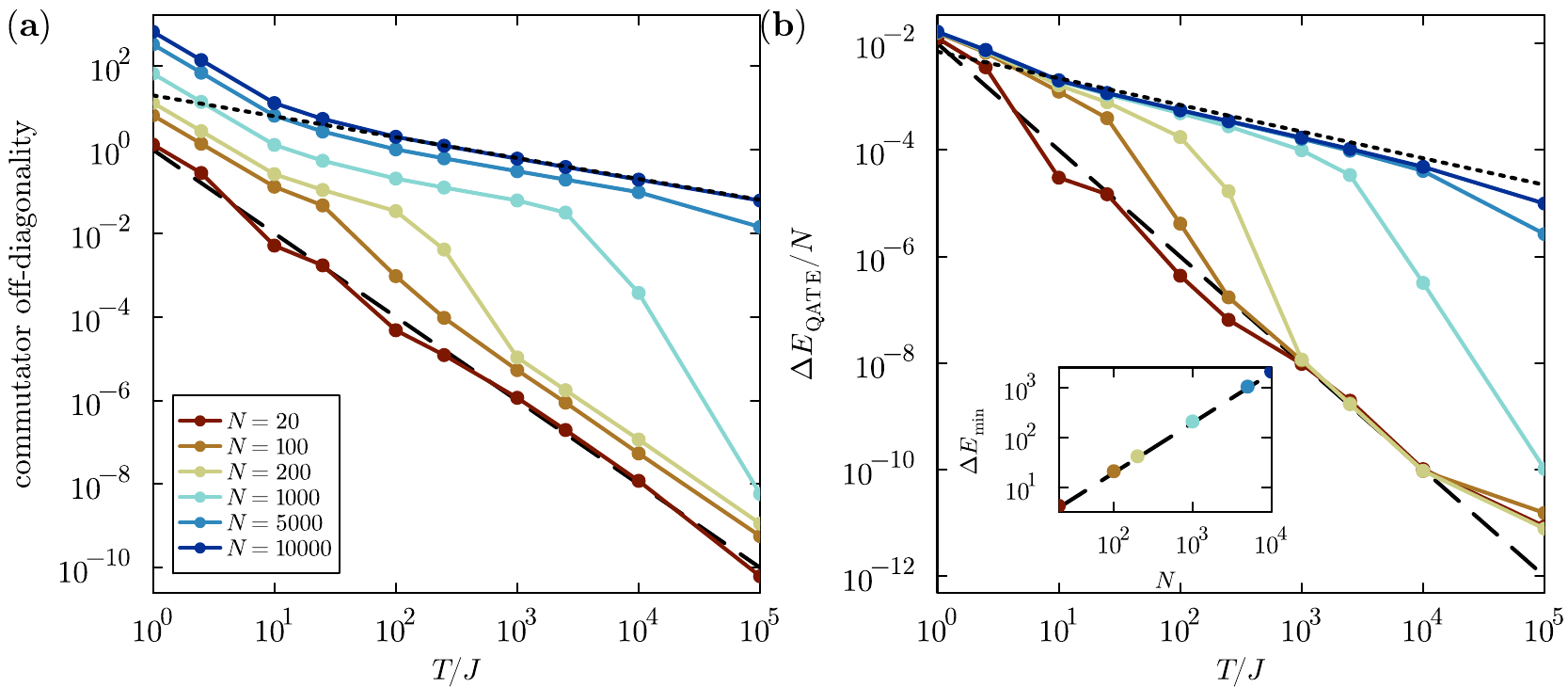}
\caption{
Effect of traversing a phase transition of QATE for TFIM, evolving from $g = 0.8 \to 1.2$, $J=1$ at $\beta = 1$.
(a) COD dependence on $T$ for various system sizes $N$, along with $T^{-2}$ and $T^{-1/2}$ trendlines (black long-dash and short-dash respectively). (b) $\dEqate/N$ dependence on $T$, along with $T^{-2}$ and $T^{-1/2}$ trendlines (black long-dash and short-dash respectively). \textit{Inset}: $\dEmin$ for different system sizes. }
\label{fig:across_transition}
\end{figure*}

We investigate the effect of crossing the phase transition at $g=1$ for the TFIM. From \cref{eq:TFI_fourier}, we see that the gap of the mode $k = -N/2$ vanishes at $g=1$. However, as discussed in \cref{sec:single_mode}, this mode remains unchanged under QATE. Thus, the closest modes with nontrivial evolution are $k = \pm (N/2-1)$. For these modes, the spectral gap is
\begin{align}
    \label{eq:clossing_gap}
    \Delta_k &= 2\epsilon_k = 2\sqrt{2+2\cos\left(\pm\frac{2\pi}{N}\left(\frac{N}{2}-1\right)\right)} \nonumber\\
    &= 2\sqrt{2-2\cos\left(\frac{2\pi}{N}\right)} \nonumber\\
    &\approx 2\sqrt{2-(2-4\pi^2/N^2)} = \frac{2\pi}{N},
\end{align}
which closes as $\mathcal{O}(N^{-1})$. In \cref{fig:across_transition}, we numerically investigate the QATE from $g=0.8$ to $g=1.2$ with a linear ramp at $\beta=1$. We observe a dual behavior where both the COD and $\dEqate/N$ transition between a $T^{-1/2}$ scaling at short times and a $T^{-2}$ scaling at long times. This transition can be explained by the QAA result \cref{eq:qaa} and the closing of the gap in \cref{eq:clossing_gap}. For times $T=\mathcal{O}(N^3)$, the adiabaticity condition is satisfied and the expected $T^{-2}$ scaling is recovered, as seen in \cref{fig:comb_tfi}. For shorter times, we observe a Kibble-Zurek-type scaling due to crossing the phase transition \cite{Kibble_1976, Kibble_1980, Zurek_2014}. In this regime, the adiabatic approximation breaks down as $T^{-1/2}$, in agreement with the theoretical prediction for gapless transitions \cite{avron1998adiabatic}.

\subsection{TFIM results with a smooth adiabatic schedule}\label{app:smooth_ramp}

\begin{figure*}[t!]
\centering
\includegraphics[width=\linewidth]{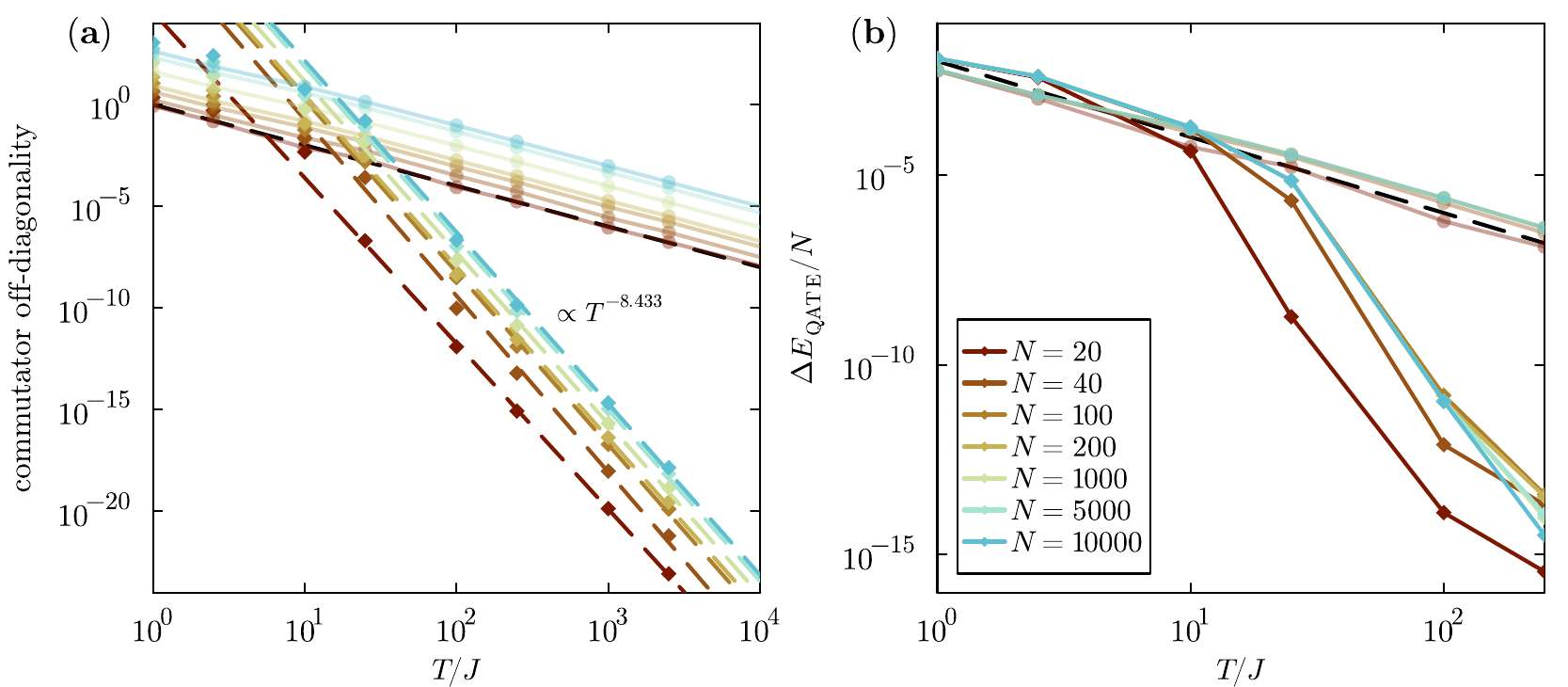}
\caption{Effect of a non-linear ramp of QATE for TFIM, evolving from $g = 1.1 \to 1.5$, $J=1$ at $\beta = 1$. (a) COD dependence on $T$ for various system sizes $N$, along with the $T^{-2}$ trendline (black dashed). Circle markers and faint lines correspond to the linear ramp and the diamond markers to the non-linear ramp. The dashed colored lines correspond to a polynomial fit of the non-linear ramp data for $T\geq100$. We observe a consistent behavior for larger systems with approximate scaling of $T^{-8.433}$ for $N=10000$. (b) $\dEqate/N$ dependence on $T$ for various $N$. Circle markers and faint lines correspond to the linear ramp and the diamond markers to the non-linear ramp. At large times $T > 10^2$, significant numerical errors accumulate.
}
\label{fig:tfi_smooth_ramp}
\end{figure*}

In \cref{sec:QAA}, we discuss how better error scaling can be achieved by using an adiabatic schedule whose derivatives vanish at the endpoints. We choose a ramp $\gamma(s)$ with the first three derivatives vanishing at $s=0$ and $s=1$:
\begin{equation}
\gamma(s) = \sin\left(\frac{\pi}{2} \sin^2\left(\frac{\pi s}{2}\right)\right)^2.
\end{equation} 
In \cref{fig:tfi_smooth_ramp}, we investigate the effect of this smooth ramp. As expected, we observe a much more rapid decay of the COD and $\dEqate$. For times beyond $T > 10^2$, we encounter numerical errors due to the extremely low values reached. Prior to this, both quantities exhibit a significantly faster decay compared to the case of a linear ramp.

\subsection{Isospectral Model at Various Temperatures}
\label{app:isospectral_betas}

\begin{figure*}[t!]
\centering
\includegraphics[width=\linewidth]{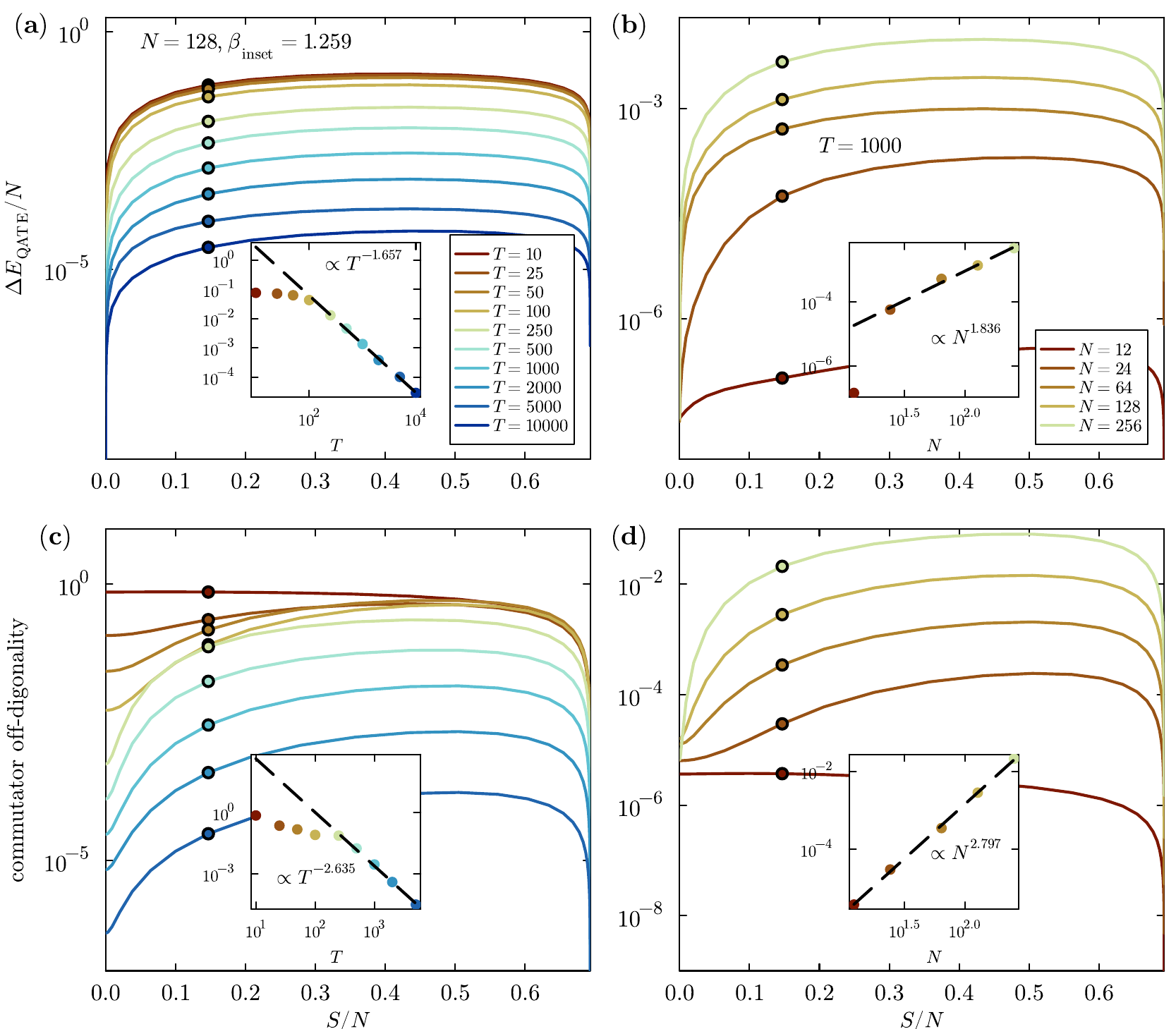}
\caption{Effect of the von Neumann entropy normalized by system size $S/N$ on QATE from the isospectral $H_{\mathrm{Z}}$ to $H_{\mathrm{TFI}}$ with $J = 1$, $g = 1.5$. 
(a) $\dEqate/N$ as a function of $S/N$ at $N = 128$ and various evolution times $T$. \textit{Inset}: scaling with evolution time $T$ fitted at selected points ($T \geq 100$, $\beta_{\mathrm{inset}} = 1.259$), using a polynomial fit (black dashed line). Circle markers in the inset correspond to the circle markers in the main plot.
(b) $\dEqate$ as a function of $S/N$ at $T = 1000$ and various system sizes $N$. \textit{Inset}: scaling with system size $N$ fitted at selected points ($\beta_{\mathrm{inset}} = 1.259$), using a polynomial fit (black dashed line). Circle markers in the inset correspond to the circle markers in the main plot.
(c) COD as a function of $S/N$ at $N = 128$ and various evolution times $T$. \textit{Inset}: scaling with evolution time $T$ fitted at selected points ($T \geq 100$, $\beta_{\mathrm{inset}} = 1.259$), using a polynomial fit (black dashed line). Circle markers in the inset correspond to the circle markers in the main plot.
(d) COD as a function of $S/N$ at $T = 1000$ and various system sizes $N$. \textit{Inset}: scaling with system size $N$ fitted at selected points ($\beta_{\mathrm{inset}} = 1.259$), using a polynomial fit (black dashed line). Circle markers in the inset correspond to the circle markers in the main plot.}
\label{fig:isospectral_betas}
\end{figure*}

In \cref{fig:isospectral_betas}, we plot the COD and $\dEqate/N$ across a range of $\beta$ values, showing their dependence on $T$ and $N$. The goal is to demonstrate that for all moderate $\beta$ values, the QATE exhibits universal behavior with consistent scaling trends. This complements the results presented in \cref{sec:results}, where we used $\beta = 1$ as a representative case. We observe irregular scaling behavior for both the COD and $\dEqate$ at low $N$ and $T$ values. In the long-time limit, however, both quantities converge to a consistent scaling regime. This allows us to extract the asymptotic scaling with respect to both $T$ and $N$. More importantly, in the large $N$ and $T$ limit, both $\dEqate$ and the COD exhibit regular behavior as functions of $S/N$, suggesting the generality of QATE performance across a wide range of temperatures.

\subsection{Ising model with mixed fields}\label{app:extra_non_int}

\begin{figure*}[t!]
    \includegraphics[width=\linewidth]{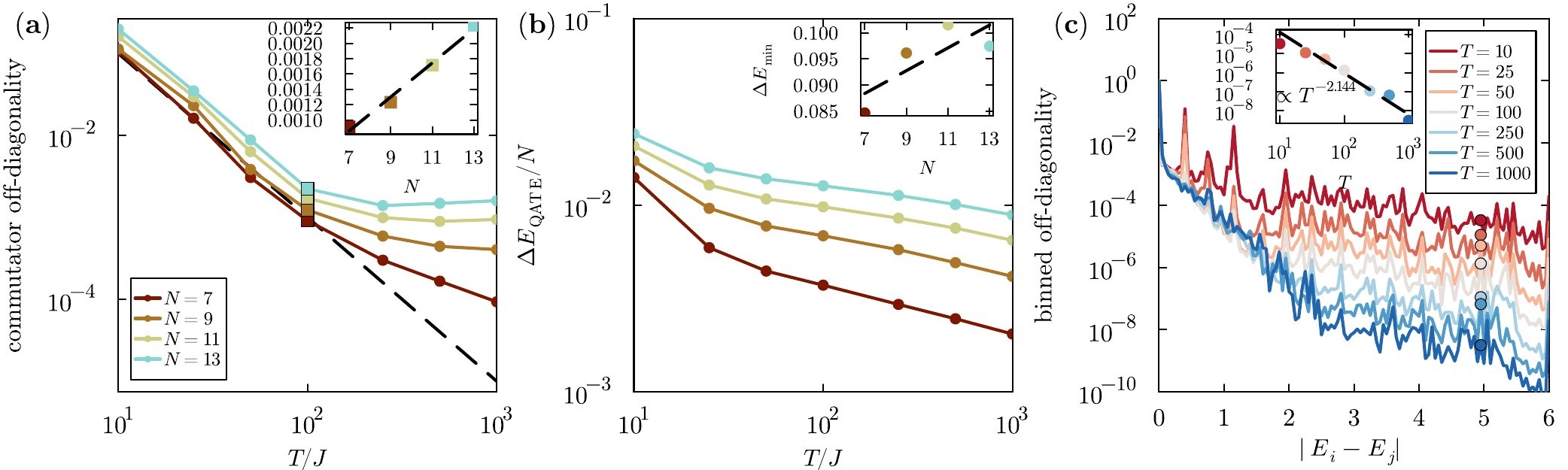}
    \caption{Ising model with mixed field evolved from $(J, h, g) = (1, 0.5, 0.0)\xrightarrow{}(1,0.5,1.05)$ at $\beta=1$ (crossing zero temperature phase transition and starting from a degenerate initial state). (a) Dependence of COD on the evolution time $T$ for various system sizes $N$, along with a $T^{-2}$ trendline (black dashed line). \textit{Inset}: COD for different system sizes at $T=100$, along with a linear fit (black dashed line). Square markers in the inset correspond to the square markers in the main plot. (b) Dependence of energy error normalized with the system size $\dEqate/N$ on $T$ for various system sizes $N$. \textit{Inset}: $\dEmin$ for different system sizes, along with a linear fit (black dashed line). (c) BOD dependence on the energy eigenvalue difference $\abs{E_i - E_j}$ for various $T$. Inset: polynomial fit of selected points (black dashed line) at the tails (circle markers in the inset correspond to the circle markers in the main plot).}
    \label{fig:ed_comm_e_filter}
\end{figure*}

To benchmark the dependence on the choice of initial state, we consider the Ising model with mixed fields (\cref{eq:ising_mixed_fields}) and initialize from a Hamiltonian $\Hi$ with parameters $(J, h, g) = (1, 0.5, 0)$. Since this Hamiltonian consists of commuting terms, it is straightforward to prepare the initial thermal state on a quantum device \cite{yimin_2016, kastoryano2016}. As the target Hamiltonian $\Hf$, we choose $(J, h, g) = (1, 0.5, 1.05)$, matching the couplings used in \cref{sec:res_non_int} and \cref{fig:ed_tfi_start}. In \cref{fig:ed_comm_e_filter}, we benchmark the performance of the QATE for this initial state. Notably, this choice of couplings causes the QATE to cross a zero-temperature phase transition. As discussed previously, crossing a phase transition can degrade performance, as observed for the TFIM (\cref{fig:comb_tfi}, \cref{fig:across_transition}).

\begin{figure*}[t!]
    \centering
    \includegraphics[width=\linewidth]{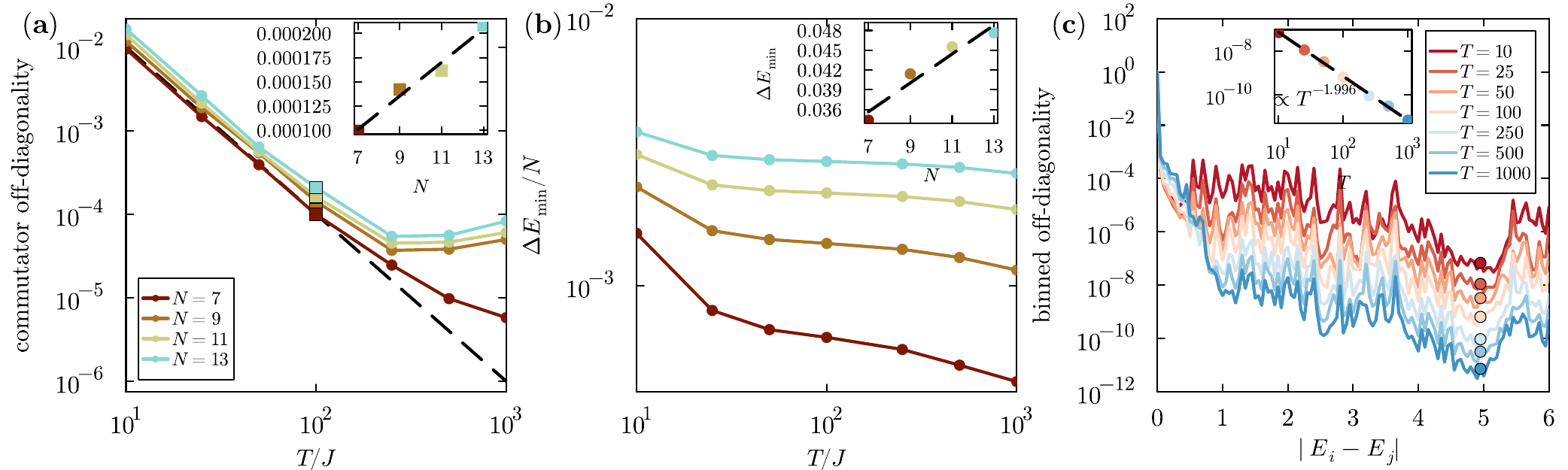}
    \caption{Ising model with mixed field evolved from $(J, h, g) = (1, 0.5, 0.0)\xrightarrow{}(1,0.5,0.5)$ at $\beta=1$ (avoiding the zero temperature phase transition and starting from a degenerate initial state). (a) Dependence of COD on the evolution time $T$ for various system sizes $N$, along with a $T^{-2}$ trendline (black dashed line). \textit{Inset}: COD for different system sizes at $T=100$, along with a linear fit (black dashed line). Square markers in the inset correspond to the square markers in the main plot. (b) Dependence of energy error normalized with the system size $\dEqate/N$ on $T$ for various system sizes $N$. \textit{Inset}: $\dEmin$ for different system sizes, along with a linear fit (black dashed line). (c) BOD dependence on the energy eigenvalue difference $\abs{E_i - E_j}$ for various $T$. Inset: polynomial fit of selected points (black dashed line) at the tails (circle markers in the inset correspond to the circle markers in the main plot).}
    \label{fig:ed_no_trans}
\end{figure*}

To investigate this further, we also consider an evolution that stops at $(J, h, g) = (1, 0.5, 0.5)$, before reaching the transition (\cref{fig:ed_no_trans}) \cite{ising_phase}. In this case, we do not observe any qualitative differences.

In \cref{fig:ed_comm_e_filter} and \cref{fig:ed_no_trans}, both the COD and BOD exhibit an approximate $T^{-2}$ scaling, consistent with the behavior observed for the integrable model and in \cref{fig:ed_tfi_start}. For moderate $T$ values, the COD scales linearly with $N$, consistent with the TFIM results but worse than in \cref{fig:ed_no_trans}. The quantity $\dEqate/N$ also converges much more slowly than the $T^{-2}$ scaling observed for the TFIM. At large $T$ values, the decay of both the COD and $\dEqate$ appears to flatten, with only marginal further improvements.

To understand this behavior, we examine the BOD, which can be separated into two regimes. In \cref{fig:ed_comm_e_filter}, for $\abs{E_i - E_j} \in [0, 2]$, we observe an exponential decay with $\abs{E_i - E_j}$, with a slope that appears independent of $T$. This $T$-independence explains the slowing of the COD decrease and the slower decay of $\dEqate$. Beyond this range, we observe a \emph{tail} regime that decays approximately as $T^{-2}$, as seen for the TFIM and in \cref{fig:ed_tfi_start}.
In \cref{fig:ed_no_trans}, the exponential decay extends over $\abs{E_i - E_j} \in [0, 1]$, suggesting a possible connection to the coupling $g$. We observe qualitative agreement with \cref{fig:ed_comm_e_filter}, confirming that crossing the zero-temperature phase transition does not significantly affect the results for this model. However, both instances considered here perform substantially worse than in \cref{fig:ed_tfi_start}.
We attribute this reduced performance to the high degeneracy of the initial state used in these cases. We conclude that the QATE performs significantly better when the initial state is non-degenerate, regardless of whether a phase transition is crossed. Better performance with a non-degenerate starting state is consistent with expectations based on the QAA.

\subsection{Temperature dependence for the Ising model with mixed fields}
\label{app:temperature_non_int}

\begin{figure*}[t!]
    \centering
    \includegraphics[width=\linewidth]{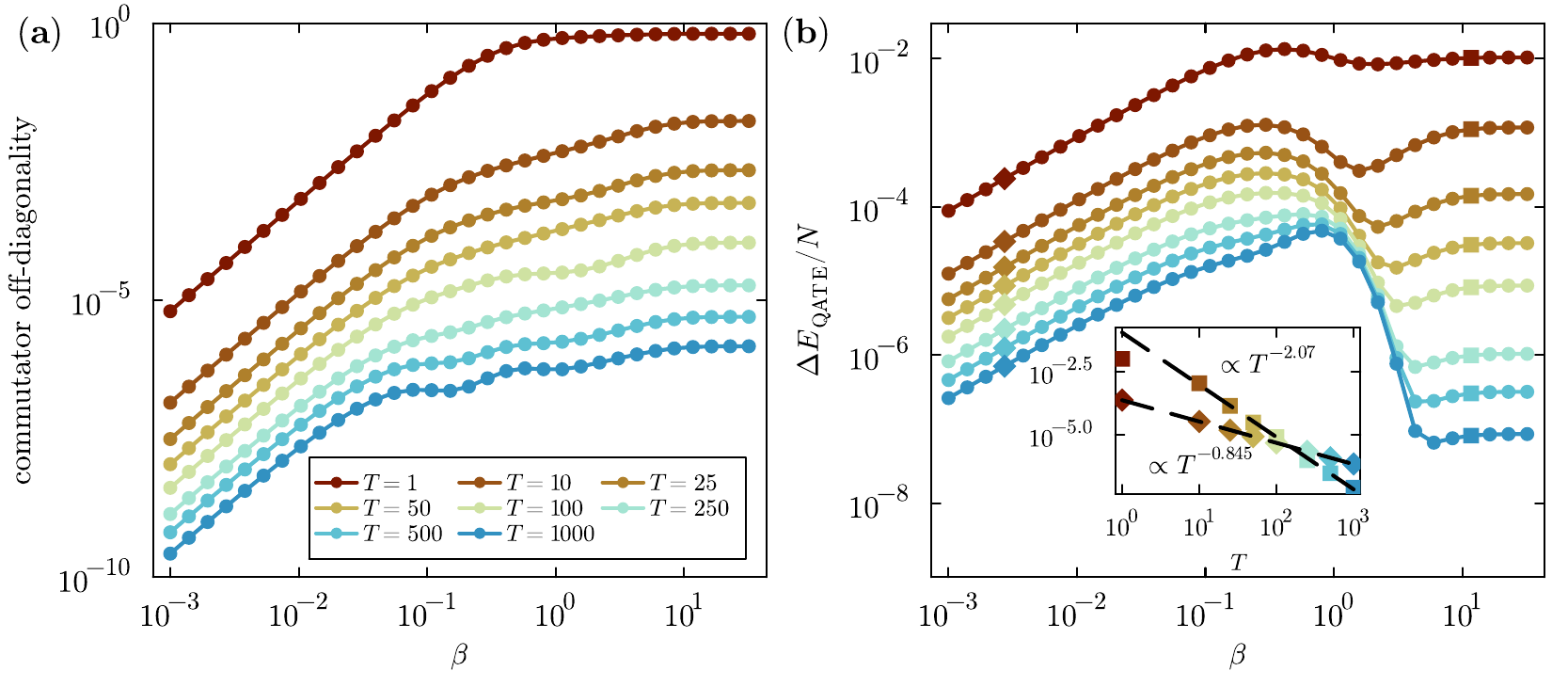}
    \caption{Ising model with mixed field evolved from $(J, h, g) = (1, 0.0, 1.05)\xrightarrow{}(1,0.5,1.05)$ at $\beta=1$, $N=13$. (a) Dependence of COD on the inverse temperature $\beta$ for various evolution times $T$. (b) Dependence of energy error normalized with the system size $\dEqate/N$ on $\beta$ for various evolution times $T$. \textit{Inset}: $\dEqate$ dependence on $T$ at $\beta\ll1$ and $\beta\gg1$, along with a linear fit for both (black dashed lines).}
    \label{fig:ed_beta_dep}
\end{figure*}

Here we investigate the temperature dependence of the Ising model with mixed fields. Specifically, we initialize the evolution in a Gibbs state of $\Hi$ with parameters $(J, h, g) = (1, 0.0, 1.05)$ at various inverse temperatures $\beta$. As the target Hamiltonian $\Hf$, we choose $(J, h, g) = (1, 0.5, 1.05)$, identical to the choice in \cref{fig:ed_tfi_start}. In \cref{fig:ed_beta_dep}, we analyze the QATE dynamics of this model for $N=13$ as a function of the inverse temperature $\beta$ and for several QATE evolution times $T$. For these parameter values, the minimum spectral gap takes the value $\Delta \approx 0.306$, at the start of the evolution.

We find that the $\dEqate$ exhibits a nonmonotonic dependence on $\beta$, with a peak at intermediate temperatures and more favorable behavior in both the low- and high-temperature limits. The worst-case performance occurs around $\beta \approx 1$, which corresponds to the parameter regime considered in the main text. Notably, near $\beta \approx 1$ the dependence of $\dEqate$ on the QATE evolution time $T$ changes qualitatively. This behavior is illustrated in the inset of \cref{fig:ed_beta_dep}(b).

For $\beta \gg 1$, we observe the expected $T^{-2}$ scaling associated with the adiabatic ground-state preparation. In this regime, ground-state behavior is expected to dominate once $\beta \Delta \gtrsim 1$, which is consistent with the observed minimal spectral gap $\Delta$ and the temperature range over which ground-state scaling is recovered. In contrast, in the high-temperature regime $\beta \ll 1$, we find a slower decay, approximately $T^{-0.85}$. Importantly, this reduced scaling appears to be systematic and persists even at very long evolution times.

\end{document}